\documentclass[a4size,hidelinks,journal]{IEEEtran}
\usepackage[english]{babel}
\usepackage{amsmath,amsfonts,amssymb, amsthm}
\usepackage{algorithmic}
\usepackage{cite} 

\newcommand{\algorithmicinitialize}{\textbf{Initialize:}}
\newcommand{\INITIALIZE}{\item[\algorithmicinitialize]}
\usepackage{algorithm}
\usepackage{array}
\usepackage[caption=true,font=footnotesize,labelfont=sf,textfont=sf]{subfig}
\usepackage{stfloats}
\usepackage{url}
\usepackage{verbatim}
\usepackage{graphicx}
\def\BibTeX{{\rm B\kern-.05em{\sc i\kern-.025em b}\kern-.08em
    T\kern-.1667em\lower.7ex\hbox{E}\kern-.125emX}}
\usepackage{balance}
\usepackage{orcidlink}
\usepackage{float}
\usepackage{lipsum}
\usepackage[acronym,shortcuts]{glossaries}
\theoremstyle{plain}

\graphicspath{{./Fig/}}

\newacronym{TX}{TX}{transmit}
\newacronym{RX}{RX}{receive}
\newacronym{IoT}{IoT}{Internet of Things}
\newacronym{SDR}{SDR}{semi-definite relaxation}
\newacronym{EVD}{EVD}{eigenvalue decomposition}
\newacronym{LDT}{LDT}{Lagrangian Dual Transform}
\newacronym{QT}{QT}{Quadratic Transform}
\newacronym{CDF}{CDF}{cumulative distribution function}
\newacronym{AP}{AP}{access point}
\newacronym{SINR}{SINR}{signal to interference-plus-noise ratio}
\newacronym{SIC}{SIC}{successive interference cancellation}
\newacronym{CSI}{CSI}{channel state information}
\newacronym{LoS}{LoS}{line-of-sight}
\newacronym{NLoS}{NLoS}{non-LoS}
\newacronym{mMIMO}{mMIMO}{massive multiple-input multiple-output}
\newacronym{MIMO}{MIMO}{multiple-input multiple-output}
\newacronym{MISO}{MISO}{multiple-input single-output}
\newacronym{SIMO}{SIMO}{single-input multiple-output}
\newacronym{SISO}{SISO}{single-input single-output}
\newacronym{MU}{MU}{multi-user}
\newacronym{JCAS}{JCAS}{joint communication and sensing}
\newacronym{JCR}{JCR}{joint communications and radar}
\newacronym{ISAC}{ISAC}{integrated sensing and communications}
\newacronym{3D}{3D}{three-dimensional}
\newacronym{2D}{2D}{two-dimensional}
\newacronym{1D}{1D}{one-dimensional}
\newacronym{ROI}{ROI}{region of interest}
\newacronym{mmWave}{mmWave}{millimeter-wave}
\newacronym{MF}{MF}{matched-filter}
\newacronym{SotA}{SotA}{state-of-the-art}
\newacronym{AWGN}{AWGN}{additive white Gaussian noise}
\newacronym{BS}{BS}{base station}
\newacronym{BF}{BF}{beamforming}
\newacronym{RF}{RF}{radio-frequency}
\newacronym{UE}{UE}{user equipment}
\newacronym{wlg}{w.l.g.}{without loss of generality}
\newacronym{CLT}{CLT}{central limit theorem}
\newacronym{PDF}{PDF}{probability density function}
\newacronym{ICI}{ICI}{inter-carrier interference}
\newacronym{BER}{BER}{bit error rate}
\newacronym{DoF}{DoF}{degrees-of-freedom}
\newacronym{VGA}{VGA}{vector Gaussian approximation}
\newacronym{FD}{FD}{full-duplex}
\newacronym{FP}{FP}{fractional programming}
\newacronym{CC}{CC}{communication-centric}
\newacronym{RC}{RC}{raised-cosine}
\newacronym{RRC}{RRC}{root raised-cosine}
\newacronym{6G}{6G}{sixth-generation}
\newacronym{V2X}{V2X}{vehicle-to-everything}
\newacronym{LEO}{LEO}{low-earth orbit}
\newacronym{I/O}{I/O}{input-output}
\newacronym{CE}{CE}{channel estimation}
\newacronym{SRM}{SRM}{sum-rate maximization}
\newacronym{ICC}{ICC}{integrated communication and computing}
\newacronym{ISCC}{ISCC}{integrated sensing, communications and computing}
\newacronym{PAM}{PAM}{pulse amplitude modulation}
\newacronym{iid}{i.i.d.}{independent and identically distributed}
\newacronym{MEC}{MEC}{mobile edge computing}
\newacronym{REMS}{REMS}{reconfigurable electromagnetic structure}
\newacronym{D-RIS}{D-RIS}{diagonal reconfigurable intelligent surface}
\newacronym{BD-RIS}{BD-RIS}{beyond diagonal reconfigurable intelligent surface}
\newacronym{RIS}{RIS}{reconfigurable intelligent surface}
\newacronym{RE}{RE}{reflective element}
\newacronym{MRT}{MRT}{maximum ratio transmission}
\newacronym{ZF}{ZF}{zero forcing}
\newacronym{SVD}{SVD}{singular value decomposition}
\newacronym{CGA}{CGA}{conjugate gradient ascent}
\newacronym{QCQP}{QCQP}{quadratic constraint quadratic programming}
\newacronym{MMSE}{MMSE}{minimum mean square error}
\newacronym{CF}{CF}{cell-free}
\newacronym{KKT}{KKT}{Karush-Kuhn-Tucker}
\newacronym{FL}{FL}{Federated learning}
\newacronym{BCD}{BCD}{block coordinate descent}
\newacronym{SCA}{SCA}{successive convex approximation}
\newacronym{RBD-RIS}{RBD-RIS}{reciprocal BD-RIS}
\newacronym{SOCP}{SOCP}{second-order cone programming}
\newacronym{SCO}{SCO}{successive convex optimization}
\newacronym{MImMO}{MImMO}{multiple-input multi-antenna user multiple-output}
\newacronym{AO}{AO}{alternating optimization}
\newacronym{pp-ADMM}{pp-ADMM}{partially proximal alternating direction method of multipliers}
\newacronym{ULA}{ULA}{uniform linear array}
\newacronym{OFDM}{OFDM}{Orthogonal Frequency Division Multiplexing}
\newacronym{UMi}{UMi}{Urban Micro}
\newacronym{3GPP}{3GPP}{3rd Generation Partnership Project}

\begin{document}



\title{Reciprocal Beyond Diagonal Reconfigurable Intelligent Surface: Distributed Scattering Matrix Design and MIMO Beamforming via Fractional Programming and Manifold Optimization}

\author{
  \IEEEauthorblockN{
~Iv\'{a}n Alexander Morales Sandoval\textsuperscript{\orcidlink{0000-0002-8601-5451}},~\IEEEmembership{Graduate Student Member,~IEEE}, \\~Marko Fidanovski\textsuperscript{\orcidlink{0009-0005-2926-1604}},~\IEEEmembership{Graduate Student Member,~IEEE},
~Hyeon Seok Rou\textsuperscript{\orcidlink{0000-0003-3483-7629}},~\IEEEmembership{Member,~IEEE},\\ ~Giuseppe Thadeu Freitas de Abreu\textsuperscript{\orcidlink{0000-0002-5018-8174}},~\IEEEmembership{Senior Member,~IEEE},
~Emil Björnson\textsuperscript{\orcidlink{0000-0002-5954-434X}},~\IEEEmembership{Fellow,~IEEE}.
  }
\thanks{I. A. M. Sandoval, M. Fidanovski, H. S. Rou, G. T. F. de Abreu are with the School of Computer Science and Engineering, Constructor University (previously Jacobs University Bremen), Campus Ring 1, 28759 Bremen, Germany (emails: \{imorales, mfidanovski, hrou, gabreu\}@constructor.university).}
\thanks{E. Björnson is with the Department of Computer Science, KTH Royal Institute of Technology, Stockholm, Sweden (email: emilbjo@kth.se).}
}

\markboth{to be submitted to the ieee transactions on wireless communications, 2026}%
{How to Use the IEEEtran \LaTeX \ Templates}

\maketitle

%

\begin{abstract}

We consider the optimization of \ac{BD-RIS}-aided \ac{MU} \ac{CF}-\ac{mMIMO} systems, where the propagation environment design achieved scattering matrix optimization is complemented by developing an efficient \ac{BS} \ac{BF} scheme that effectively exploits the latter ``engineered'' channel.
In particular, we describe a \ac{FP} method, which based on the equivalent channel incorporating a reciprocal \ac{BD-RIS} (RBD-RIS) parameterized by existing scattering matrix design methods, yielding the correspondingly optimized \ac{MIMO} \ac{BF} weights.
%
The proposed approach decomposes the \ac{TX} beamformer into multiple \ac{SRM} sub-beamformers, each satisfying an independent power-constraint, such that distributed \ac{MIMO}-\ac{BF} scenarios can be optimally handled.
Although the proposed \ac{SRM}-\ac{MIMO}-\ac{BF} framework is independent of the specific scattering matrix design, extending the \ac{BD-RIS}-aided system model to the \ac{CF}-\ac{mMIMO} setting requires the design of a corresponding beamforming matrix. 
In this context, this work investigates the impact of beamforming in \ac{RIS}-aided systems.
Simulation results demonstrate that the proposed method for designing the \ac{MIMO}-\ac{BF} weights, when combined with the previously developed design of reciprocal \ac{BD-RIS} (RBD-RIS) scattering matrices, outperforms existing \ac{BD-RIS}-aided \ac{SotA} schemes employing existing \ac{MIMO}-\ac{BF} techniques, indicating that the whole contribution is more than the sum of the parts.
\end{abstract}



\glsresetall
\glslocalunset{BF}

\vspace{-2ex}
\section{Introduction}

\IEEEPARstart{B}{eamforming} (BF) is an essential technique employed at either or both ends of modern wireless communications systems, which utilizes knowledge of the channel state to mitigate interference, enhance signal quality, add robustness, increase achievable rates, and improve the energy efficiency of the system.
In turn, \acp{RIS} are deployed to the environment itself, having similar effects as \ac{BF}, but without requiring significant power and \ac{RF} processing such as down-conversion, sampling, decoding, etc.

Unlike \ac{BF}, which reacts to a given channel state, \acp{RIS} fundamentally modify the propagation channel itself by selectively reflecting impinging waves \cite{BjornsonSPM2022, LiuCST2021, NeriniTWC2023}.
As such, \ac{BF} and \ac{RIS} have synergistic roles, such that the optimization of \ac{BF} weights and \ac{RIS} parameters needs to be considered jointly.

In our prior work on this topic~\cite{FidanovskiArx2025,FidanovskiTWC2026} a novel method to design \ac{RBD-RIS} scattering matrices was proposed, whereby the challenging problem of maximizing the system's sum-rate while incorporating a reciprocal structure that enforces both unitary and symmetry constraints was addressed under a manifold optimization framework.
In accordance with \ac{SotA} approaches, the system model considered in the aforementioned work was a \ac{MU}-\ac{MISO} setup, which facilitates comparison, analysis, and highlights the gains obtained from the scattering matrix design.
Nevertheless, the assumption of a single antenna user is restrictive, thereby motivating the extension to a generalized \ac{MU} \ac{CF}-\ac{mMIMO} setup.

The proposed solution provides a structured approach applicable to various \ac{BD-RIS} architectures, namely single-, group-, and fully-connected, while also showing promising performance improvements compared to the \ac{SotA} methods.
Building upon the foundation established in \cite{FidanovskiTWC2026}, this article addresses the accompanying system optimization problem, namely, the beamforming matrix design.
To elaborate further, while the scattering matrix determines how signals are manipulated at the \ac{RIS}, the beamforming matrix specifies the active transmission strategy at the \acp{AP} \cite{BjornsonSPM2024, GershmanSPM2010, GodaraCRCP2018}, such that the joint optimization of both  is crucial to fully exploit the potential of \ac{BD-RIS} structures.

In recognition of the above, considerable research has been done on the design of beamforming methods in combination with \acp{RIS} \cite{LiTSP2024, LiJSAC2023,LiuWCL2024, ChenSPAWC2024, ZhouTWC2024, LiTWC2022}. 
Among such approaches, a widely adopted strategy for \ac{TX}-\ac{BF} design is the use of \ac{FP} techniques, which effectively handle non-convex optimization problems and provide near-optimal solutions, making them well-suited to systems seeking high performance with manageable computational complexity.
To cite an excellent example, Li $et$ $al.$ \cite{LiTSP2024} formulated two beamforming design problems for \ac{BD-RIS} with reflective and hybrid/multi-sector modes.
In the first scenario, a point-to-point \ac{MIMO} system aided by a reflective mode \ac{BD-RIS} was considered, where a joint optimization of the scattering and precoding matrices was developed and addressed using a two-stage approach.
The proposed two-stage approach first designed the \ac{BD-RIS} scattering matrix to maximize the effective channel strength, and then obtained the precoding matrix from the left unitary matrix following an \ac{SVD} of the resulting channel.
In the second scenario, a \ac{MU}-\ac{MISO} system aided by a hybrid/multi-sector mode \ac{BD-RIS} was considered, and a similar joint optimization problem was proposed.
The problem was then reformulated using \ac{FP} techniques, and transformed into a four-block optimization problem, where the blocks were iteratively updated until convergence.
Namely, the precoding block solution was obtained based on the \ac{KKT} conditions.
A similar solution, based on the \ac{FP} approach for the \ac{TX} beamformer, was obtained in \cite{LiJSAC2023}.
Both works are limited in that the scattering matrix design does not account for the reciprocal structure of the \ac{BD-RIS}.

Beamforming designs for \acp{RIS} have also been investigated in the context of \ac{ISAC} systems \cite{LiuWCL2024, ChenSPAWC2024}.
For instance, in \cite{LiuWCL2024}, a \ac{MU}-\ac{MISO} \ac{ISAC} system aided by a fully-connected \ac{BD-RIS} was considered, where a joint optimization problem was formulated for the design of the linear filter, beamforming and scattering matrices, aiming to maximize network throughput while satisfying radar sensing quality constraints.
The problem was then addressed using \ac{FP} techniques combined with the majorization-minimization method, transforming it into a convex form and enabling computation of the optimal \ac{TX} beamformer. 
Additionally, in \cite{ChenSPAWC2024}, a \ac{mmWave} \ac{ISAC} system with a \ac{BD-RIS}-aided transmitter was studied, where a joint design of digital beamforming, scattering, and digital radar covariance matrices was proposed to enhance both communication and sensing performance.
Following a conventional two-stage approach, the scattering matrix was optimized first, followed by the \ac{TX} beamformer.
The corresponding \ac{TX} beamforming matrix was obtained by formulating an optimization problem, introducing slack variables, and applying multiple \ac{SCA} iterations along with an \ac{EVD} on the resulting suboptimal solution.

An alternative approach was proposed in \cite{ZhouTWC2024}, where a power minimization and energy efficiency maximization problem was formulated for a downlink \ac{MU}-\ac{MISO} network, each requiring the design of the corresponding \ac{TX} beamformers.
For power minimization, the \ac{AP} beamformer is obtained via a \ac{SOCP} problem formulation, allowing for an optimal solution to be obtained using standard numerical solvers like CVX \cite{GrantCVX2014}.
In turn, for the energy efficiency maximization problem, the \ac{AP}-\ac{BF} matrix was designed following conventional \ac{FP} transformation methods.
While effective, the approach relies on generic convex optimization solvers and entails relatively high computational complexity.
To alleviate this issue, a related approach was presented in \cite{WuArx2024}, where the beamforming matrix was optimized as a block variable using a \ac{pp-ADMM} framework, enabling more efficient iterative updates.

Additionally, in \cite{LiTWC2022} a \ac{RIS}-aided \ac{MU}-\ac{MISO} system was considered and an optimization problem was formulated.
Similarly to the approach in this article, \ac{FP} techniques are applied to the objective function, followed by a Lagrange multiplier method, allowing for the optimal precoder to be obtained by verifying the first-order optimality conditions.
The Lagrange multiplier was determined using a simple bisection search.


A common limitation of all the aforementioned methods\footnote{This limitation applies also to alternative \ac{BF} that are not based on the \ac{FP} approach, e.g., \cite{FangCL2024, XuWCNC2025}.} is, however, that the beamforming matrix is obtained without enforcing sub-array power constraints, such that the results cannot be used directly in distributed \ac{BF} scenarios.
To elaborate further, if not a single but a set of \acp{AP} collaborate in a distributed \ac{BF} fashion to transmit signals to users \cite{BjornsonDSP2019}, under a condition where each \ac{AP} is subjected to its own \acp{TX} power constraint\cite{MaTWC2023, LiuJSTSP2025,KatsanosArx2026}, $e.g.$, due to particular hardware limitations or local regulatory restrictions \cite{ShiIWCMC2016,GhaniPONE2016},\cite{PearceER2020, RodaESCP2014}, the optimal collective \ac{BF} matrix must have blocks with distinct powers.
This constraint is difficult to enforce under conventional approaches where the entries of the \ac{BF} matrix are optimized as free-complex numbers, that is, such that both the amplitude and the phase of antenna coefficients are allowed to take on any values.

In this context, \cite{MaTWC2023} considered a downlink \ac{RIS}-aided \ac{CF}-\ac{mMIMO} network, whereby a \ac{FP} approach along with \ac{AO} methods were utilized to obtain the hybrid \ac{BF} matrix at the \ac{BS} under per-\ac{AP} power constraints.
This work is however limited in that it only considers a conventional diagonal \ac{RIS} structure.
In a related direction, \cite{KatsanosArx2026} proposed a decentralized \ac{BF} framework for multi-\ac{BD-RIS}-assisted \ac{CF}-\ac{mMIMO} \ac{OFDM} systems, where the non-convex sum-rate maximization problem is addressed via successive convex approximation using first-order surrogate functions. 
In particular, that work considers a wideband setting with multiple shared \acp{BD-RIS} and jointly optimizes the precoders together with the tunable capacitance and permutation matrices through consensus-based updates. 
Rather than designing the scattering matrix directly, the passive configuration is obtained through a circuit-based parametrization that guarantees reciprocity. 
This offers a valuable and practically motivated perspective on scalable multi-\ac{BD-RIS} optimization.

In contrast to prior approaches, this article addresses the joint design of per-\ac{AP} \ac{TX} \ac{BF} and multiple \ac{RBD-RIS} scattering matrices using \ac{FP} techniques, which transform the original non-convex problem into a sequence of tractable convex subproblems, enabling efficient optimization under practical constraints.
To the best of our knowledge, this is the first work to consider the aforementioned \ac{FP}-framework in a general \ac{CF}-\ac{mMIMO} setting.
Overall, the main contributions of this article can be summarized as follows:
\begin{itemize}
    \item A novel \ac{BF} matrix design is proposed for \ac{RBD-RIS}-aided \ac{MU} \ac{CF}-\ac{mMIMO} systems, where the amplitude and phase coefficients of the antenna elements are optimized separately, offering flexibility that enables distributed (collaborative) \ac{BF}.
    \item A generalized closed form expression for the gradient of the objective function with respect to the scattering matrix is provided, enabling its optimization using the algorithm proposed in{\cite{FidanovskiTWC2026}} over multiple independent \ac{RBD-RIS}.
    \item A closed form expression for the proposed \ac{FP}-based sub-\ac{BF} matrix is given, which allows for an optimal solution to be found using a simple \ac{2D}-search, yielding overall a low-complexity alternative to \ac{BF} for \ac{RBD-RIS}-aided \ac{MU} \ac{CF}-\ac{mMIMO} systems.
\end{itemize}

\noindent\textit{Organization:} The remainder of the article is organized as follows. 
Section \ref{sec:sysmodel} introduces the \ac{BD-RIS}-aided \ac{MU} \ac{CF}-\ac{mMIMO} system model, presents the decomposition of the \ac{TX} beamforming matrix, and formulates the problem considered in this article. 
Following the problem formulation, Section~\ref{sec:bfdesign} provides a solution for the \ac{TX} beamforming matrix design.
Simulation results are provided in Section~\ref{sec:simresults}, which help evaluate the performance of the proposed designs.
Conclusions and possible future work directions are discussed in Section~\ref{sec:conc}.

\noindent\textit{Notation:} Unless otherwise specified, $\mathbf{X}$ and $\mathbf{x}$ denote matrices and vectors. 
The absolute value, $\ell^2$ and Frobenius norm are denoted by $|\cdot|$,  $\|\cdot\|_2$, $\|\cdot\|_F$. 
The transpose and the Hermitian transpose are denoted by $(\cdot)^T$ and $(\cdot)^H$. $\mathbf{X}_{i,j}$ denotes the $i$-th row and $j$-th column element of the matrix $\mathbf{X}$, $\mathbf{X}_{i:\bar{i},j:\bar{j}}$ extracts the elements from the $i$-th to the $\bar{i}$-th row, and $j$-th to $\bar{j}$ column from the matrix $\mathbf{X}$, $\text{diag}(\mathbf{x})=(x_1,x_2,\dots,x_k)$ denotes a diagonal matrix, where the main diagonal is $\mathbf{x}$, $\text{blkdiag}\left(\mathbf{X}_1, \mathbf{X}_2, \dots, \mathbf{X}_k\right)$ denotes a $G\times G$ block diagonal matrix with off diagonal elements $0$, where the blocks are $\mathbf{X}_1, \mathbf{X}_2, \dots, \mathbf{X}_k$.
The set of complex and real numbers is denoted by  $\mathbb{C}$ and $\mathbb{R}$, $\Re\{\cdot\}$ and $\Im\{\cdot\}$ denote the real and imaginary values of a complex number, respectively.
$\mathcal{CN}(0, \sigma^2)$ denotes a complex normal random variable with a zero mean and $\sigma^2$ variance.
Finally, $\log_n\big|\cdot\big|$ denotes the log determinant.

\vspace{-2ex}
\section{System Model}
\label{sec:sysmodel}
\vspace{-0.5ex}

\subsection{System and Channel Model}

\vspace{-0.5ex}

\begin{figure}[H]
  \centering  
  \includegraphics[width=1\linewidth]{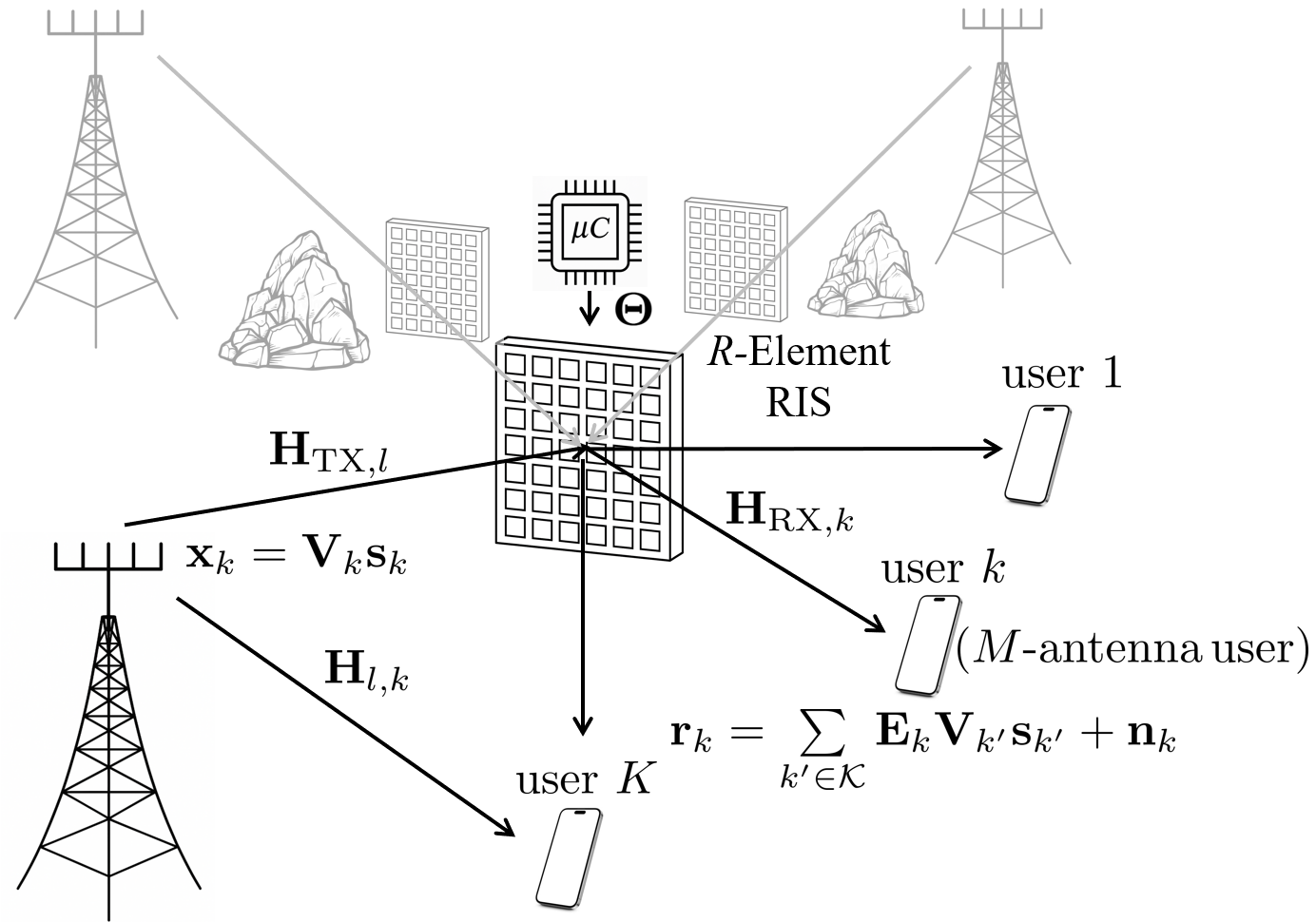}
  \caption{{Illustration of the system model, where $L$ \acp{AP} with $N_a$ \ac{TX} antennas each serve $K$ users with $M$ \ac{RX} antennas each through multiple $R$-element \acp{RBD-RIS} with a direct \ac{LoS} link between the \acp{AP} and the users.\protect\footnotemark}}
  \label{fig:system model}
  \vspace{-1ex}
\end{figure}

Consider a downlink \ac{BD-RIS}-aided \ac{MU} \ac{CF}-\ac{mMIMO} system as illustrated in Figure \ref{fig:system model}, where $L$ interconnected \acp{AP} with $N_a$ \ac{TX} antennas each, {resulting in a} total number of $N_t = L \times N_a$ {antennas}, service $K$ users with $M$ \ac{RX} antennas each, {with the assistance of a \ac{BD-RIS}} consisting of {$R$} \acp{RE} located between the \acp{AP} and users{\footnote{{The extension to multiple \ac{BD-RIS} is detailed in Appendix \ref{sec:appendix}.}}}.

For convenience, let $\mathcal{L}\triangleq \{1,2,\dots,L\}$ and $\mathcal{K}\triangleq\{1,2,\dots,K\}$ be the sets of indices $l$ and $k$, corresponding to \acp{AP} and users, respectively.
Furthermore, $\mathbf{\Theta}\in \mathbb{C}^{{R\times R}}$ denotes the \ac{BD-RIS} scattering matrix, the channel linking the $l$-th \ac{AP} and the \ac{BD-RIS} (i.e., the \ac{AP}$_l${-to}-\ac{BD-RIS} channel) is denoted as $\mathbf{H}_{\mathrm{TX},l}\in \mathbb{C}^{{R}\times N_a}$, and the \ac{BD-RIS}-{to-user $k$} channel matrix is given by $\mathbf{H}_{\mathrm{RX},k} \in \mathbb{C}^{M\times {R}}$.
The transmit signal vector is denoted as $\mathbf{x}_k = \mathbf{V}_k\mathbf{s}_k$, where the information symbols $\mathbf{s}_k\in \mathbb{C}^{M\times 1}$ corresponding to {the $k$-th user} satisfy $\mathbb{E}[\mathbf{s}_k\mathbf{s}_k^H] = \mathbf{I}$, and the \ac{TX} beamforming matrix from all \acp{AP} to $U_k$, i.e., $\mathbf{V}_k\in \mathbb{C}^{N_t\times M}$ meets the power constraint $\sum_{k\in \mathcal{K}}\|\mathbf{V}_k\|_F^2\leq P_{\mathrm{max}}$, with $P_{\mathrm{max}}$ being the maximum power that is available to the system.

{The equivalent channel linking the $l$-th \ac{AP} to the $k$-th user, including both the direct \ac{LoS} and \ac{BD-RIS}-aided path can be defined as $\mathbf{E}_{l,k} = \mathbf{H}_{l,k} + \mathbf{H}_{\mathrm{RX},k}\mathbf{\Theta}\mathbf{H}_{\mathrm{TX},l}$, such that $\mathbf{E}_{l,k} \in \mathbb{C}^{M\times N_a}$.

\footnotetext{{This setting represents the most general scenario, aimed at highlighting the impact of the joint scattering and the \ac{BF} matrix design.}}

Straightforwardly, the equivalent downlink channel matrix from all \acp{AP} to $U_k$ is obtained as the concatenation of the \ac{AP}$_l$-$U_k$ equivalent channels, namely
\begin{equation}
\label{eq:equivalent_channel}
    \mathbf{E}_k \triangleq [\mathbf{E}_{1,k},\mathbf{E}_{2,k},\dots, \mathbf{E}_{L,k}] \in \mathbb{C}^{M\times N_t}.
\end{equation}}
\vspace{-2ex}

{Considering the channel model, the matrices $\mathbf{H}_{\mathrm{TX},l}$, $\mathbf{H}_{\mathrm{RX},k}$, and $\mathbf{H}_{l,k}$ are generated using a narrowband Rician fading model with distance-dependent large-scale fading.\footnote{{Details on the large-scale fading model are provided in Section~\ref{sec:simresults}, where different parameters are specified for the \ac{AP}-to-\ac{BD-RIS}, \ac{BD-RIS}-to-user, and direct \ac{AP}-to-user links.}}
Specifically, each channel realization is expressed as
\begin{equation}
\mathbf{H} = \sqrt{\Upsilon(d)} \left( \sqrt{\frac{K_f}{K_f+1}} \mathbf{H}_{\mathrm{LoS}} + \sqrt{\frac{1}{K_f+1}} \mathbf{H}_{\mathrm{NLoS}} \right),
\end{equation}
where $\Upsilon(d)$ denotes the large-scale fading coefficient and $K_f$ is the Rician factor. 
The NLoS component $\mathbf{H}_{\mathrm{NLoS}}$ consists of independent and identically distributed complex Gaussian entries, whereas the LoS component is modeled using a geometric rank-one representation based on \acp{ULA}, given by
\begin{equation}
\mathbf{H}_{\mathrm{LoS}} = e^{-j \frac{2\pi d}{\lambda}} \mathbf{a}_{\mathrm{rx}}(\theta_{\mathrm{AoA}})\mathbf{a}_{\mathrm{tx}}^{H}(\theta_{\mathrm{AoD}}),
\end{equation}
where $d$ is the link distance, $\lambda$ is the carrier wavelength, and $\mathbf{a}_{\mathrm{rx}}(\cdot)$ and $\mathbf{a}_{\mathrm{tx}}(\cdot)$ denote the receive and transmit array response vectors, respectively. 
The angles of arrival and departure are independently drawn from $[-\pi/2, \pi/2]$.
}

Thus, the complex baseband received signal at {the $k$-th user} can be expressed as
\begin{equation}
  \mathbf{r}_k = \overbrace{\,\mathbf{E}_k \mathbf{V}_k \mathbf{s}_k}^{\text{Intended signal}}\; + \!\!\!\!\!\overbrace{\!\!\!\!\!\sum_{k' \in \mathcal{K}\backslash \{k\}} \!\!\!\!\!\! \mathbf{E}_k \mathbf{V}_{k'} \mathbf{s}_{k'}}^{\text{Downlink inter-user interference}} \!\! +  \,\mathbf{n}_k,
  \end{equation}
where $\mathbf{r}_k \in \mathbb{C}^{M\times 1}$ and $\mathbf{n}_k \in \mathbb{C}^{M\times 1} \sim \mathcal{CN}(0, N_0\mathbf{I})$ denotes the circularly symmetric \ac{AWGN} noise at {user $k$} with the power spectral density $N_0$.

\subsection{Problem Formulation}

From the received signal model, it follows that the 
%
{interference-plus-noise-whitened signal covariance matrix received at the $k$-th user, $\mathbf{\Gamma}_k$,} can be formulated as
\begin{subequations}
\begin{equation}
\label{eq:SINR}
    \mathbf{\Gamma}_k = \mathbf{V}_k^\mathrm{H} \mathbf{E}_k^\mathrm{H} \mathbf{\Psi}_k^{-1} \mathbf{E}_k \mathbf{V}_k ,
\end{equation}
where
\begin{equation}
\label{eq:interf}
    \mathbf{\Psi}_k \triangleq  \sum\limits_{k'\in \mathcal{K}\backslash \{k\}} \mathbf{E}_{k} \mathbf{V}_{k'} \mathbf{V}_{k'}^\mathrm{H} \mathbf{E}^\mathrm{H}_{k} + N_0  \mathbf{I}_M,
\end{equation}
is the {corresponding }interference-plus-noise {covariance matrix}.
\end{subequations}

The \ac{TX} beamformer for {user $k$} can be further divided into 
\begin{equation}
\label{eq:vk_from_qlk}
    \mathbf{V}_k \triangleq \begin{bmatrix}
                    \mathbf{Q}_{1,k} \\
                    \mathbf{Q}_{2,k} \\
                    \vdots \\
                    \mathbf{Q}_{L,k}
                    \end{bmatrix}, \forall \; k \in K,
\end{equation}
where $\mathbf{Q}_{l,k} \in \mathbb{C}^{N_a \times M}$ represents the  sub-\ac{BF} matrices from the $l$-th AP, to service {user $k$}, and can be collected into
\begin{equation}
    \mathbf{Q}_l \triangleq \Big[ \mathbf{Q}_{l,1}, \mathbf{Q}_{l,2}, \cdots \mathbf{Q}_{l,K} \Big], \forall \; l \in \mathcal{L},
\end{equation}
such that $\mathbf{Q}_l \in \mathbb{C}^{N_a \times (K \cdot M)}$ is the \ac{TX} beamforming matrix local to the $l$-th AP.

Following the formulation in{\cite{FidanovskiTWC2026}}, the power-constrained sum-rate maximization problem is described as
\begin{subequations} \label{prob:P1}
    \begin{eqnarray}
    \text{(P1)}: \quad \underset{\mathbf{V}_{k}, \mathbf{\Theta}}{\mathrm{maximize}} && \sum_{k\in\mathcal{K}} \log_2 \big|\mathbf{I}_{M} + \mathbf{\Gamma}_k \big| \label{eq:obj_P1} \\ 
    \text{subject to} &&  \sum_{k\in\mathcal{K}}\|\mathbf{V}_k\|^2_\mathrm{F} \leq P_{\max} ,\label{eq:p1_bf} \\
    && \|\mathbf{Q}_l\|^2_\mathrm{F} \leq P_{l}, \; \forall l \in \mathcal{L},\label{eq:p1_sbf} \\
    && \mathbf{\Theta} \in \mathcal{S}_{a_1}, \label{eq:p1_sm1}  \\
    && \mathbf{\Theta} \in \mathcal{S}_{a_2}, \label{eq:p1_sm2}
    \end{eqnarray}
  \end{subequations}
where $ a \in \{ \text{SC}, \text{FC}, \text{GC} \}$ {denotes the connectivity level of the \ac{BD-RIS}},  $P_\text{AP}$ is the maximum transmit power available at each AP, {such that} {$P_{\max} \triangleq \sum_{l\in\mathcal{L}} P_l$}. 

It is {evident} that the optimization problem described in \eqref{prob:P1} is inherently non-convex, due to {both} the objective function in \eqref{eq:obj_P1} {and the} constraints in \eqref{eq:p1_bf}-\eqref{eq:p1_sm2}.
{The scattering matrix is subject to \ac{DoF} restrictions imposed by the reciprocal and lossless properties of the \ac{BD-RIS}, which are generally non-convex \cite{FidanovskiTWC2026}.}
Additionally, the constraints on the beamforming and  sub-\ac{BF} matrix are non-convex.
For simplicity, the sum-rate over all users is denoted as
\begin{equation}
  {\eta} \triangleq \sum_k {\eta}_k, \quad \text{with} \quad {\eta}_k \triangleq \log_2\big|\mathbf{I}_M + \mathbf{\Gamma}_k\big|.
\end{equation}

{While \cite{FidanovskiTWC2026} focuses on the design of the scattering matrix for \ac{RBD-RIS} under specific conditions that highlight its impact on the system sum-rate, this work generalizes that study by additionally considering the second stage, namely the design of the beamforming matrix.}

\section{Stage 2: Beamforming Matrix Design}
\label{sec:bfdesign}

Since a \ac{CF}-\ac{mMIMO} system is considered in this {article}, a generalization of the solution { for the scattering matrix design proposed in \cite{FidanovskiTWC2026} }to the \ac{MIMO} case is required. 
{Straightforwardly, we mitigate the issue of non-convexity of the log-ratio term in the objective function \eqref{eq:obj_P1} by applying the matrix \ac{FP} method described in \cite{shen2018phd, ShenTSP2018, Shen2TSP2018}.

To this extent, upon applying the \ac{LDT} to ${\eta}_k$ the following is obtained
\begin{align}
    \bar{{\eta}}_k & = \log_2 \big|\mathbf{I}_M + \mathbf{Z}_k\big|\ - \mathrm{Tr}(\mathbf{Z}_k)\\
    & \hspace{2ex} + \mathrm{Tr}\Big((\mathbf{I}_M+\mathbf{Z}_k)\mathbf{V}_k^H\mathbf{E}_k^H\mathbf{\Psi}_k^{-1}\mathbf{E}_k\mathbf{V}_k \Big), \nonumber
\end{align}
where $\mathbf{\Psi}_k$ denotes the interference term at the $k$-th user defined in \eqref{eq:interf}, and the auxiliary variable (i.e., the matrix Lagrange multiplier $\mathbf{Z}_k$) is given by
\begin{equation}
\label{eq:z_k}
    \mathbf{Z}_k = \mathbf{V}_k^H\mathbf{E}_k^H\mathbf{\Psi}_k^{-1}\mathbf{E}_k\mathbf{V}_k.
\end{equation}
Furthermore, the \ac{QT} is applied due to the ratio on the variable $\mathbf{V}_k$ embedded in the last term of $\bar{{\eta}}_k$ as follows
\begin{align}
    \breve{{\eta}}_k &= \log_2 \big| \mathbf{I}_{M} + \mathbf{Z}_k \big| - \mathrm{Tr}\big( \mathbf{Z}_k \big) \\
    & \hspace{2ex} + \mathrm{Tr}\Big( \big( \mathbf{I}_{M}+\mathbf{Z}_k \big) \big(  2 \Re \{  \mathbf{V}_{k}^\mathrm{H} \mathbf{E}^\mathrm{H}_{k} \mathbf{Y}_k \}- \mathbf{Y}_k^\mathrm{H} \mathbf{\Psi}_k \mathbf{Y}_k \big) \Big), \nonumber
\end{align}
where the auxiliary variable corresponding to the \ac{QT}, namely $\mathbf{Y}_k$ is given by
\begin{equation}
\label{eq:y_k}
    \mathbf{Y}_k = \big(\mathbf{E}_k\mathbf{V}_k\mathbf{V}_k^H\mathbf{E}_k^H+\mathbf{\Psi}_k\big)^{-1}\mathbf{E}_k\mathbf{V}_k.
\end{equation}

For brevity, the constant terms are omitted and $\breve{{\eta}}_k$ is rewritten as seen in \eqref{eq:fulleta}, located at the top of the next page.

\begin{figure*}[t]
\begin{align}
\label{eq:fulleta}
    \breve{\eta}_k \! & = \mathrm{Tr}\Big( \big( \mathbf{I}_{M}+\mathbf{Z}_k \big)  \big( 2 \Re \{  \mathbf{V}_{k}^\mathrm{H} \mathbf{E}^\mathrm{H}_{k} \mathbf{Y}_k \} - \sum\limits_{k'\in \mathcal{K}} \mathbf{Y}_k^\mathrm{H} \mathbf{E}_{k} \mathbf{V}_{k'} \mathbf{V}_{k'}^\mathrm{H} \mathbf{E}^\mathrm{H}_{k} \mathbf{Y}_k \big) \Big) \\
    & = 2 \cdot \mathrm{Tr} \Big( \big( \mathbf{I}_{M}+\mathbf{Z}_k \big) \Re \{  \mathbf{V}_{k}^\mathrm{H} \mathbf{E}^\mathrm{H}_{k} \mathbf{Y}_k \} \Big) - \mathrm{Tr} \Big(\sum\limits_{k'\in \mathcal{K}} \big( \mathbf{I}_{M}+\mathbf{Z}_k \big) \mathbf{Y}_k^\mathrm{H} \mathbf{E}_{k} \mathbf{V}_{k'} \mathbf{V}_{k'}^\mathrm{H} \mathbf{E}^\mathrm{H}_{k} \mathbf{Y}_k \Big) \nonumber
\end{align}
\begin{center}
\rule{0.5\textwidth}{0.1pt}
\end{center}
\end{figure*}

To derive a general solution, the group-wise definition for the equivalent channel is considered, namely
\begin{equation}
    \mathbf{E}_{l,k}^{(g)} = \mathbf{H}_{l,k} + \mathbf{H}_{\mathrm{RX},k}^{(g)}\mathbf{\Theta}_g\mathbf{H}_{\mathrm{TX},l}^{(g)},
\end{equation}
where $\mathbf{H}_{\mathrm{RX},k}^{(g)} = \mathbf{H}_{\mathrm{RX},k[1:M, R_G(g-1) + 1:gR_G ]}$, s.t. $\mathbf{H}_{\mathrm{RX},k}^{(g)} \in \mathbb{C}^{M\times R_G}$, $\mathbf{\Theta}_g = \mathbf{\Theta}_{[R_G(g-1) + 1:gR_G, R_G(g-1)+1:gR_G]}$, s.t. $\mathbf{\Theta}_g \in \mathbb{C}^{R_G \times R_G}$, $\mathbf{H}_{\mathrm{TX},l}^{(g)} = \mathbf{H}_{\mathrm{TX},l[R_G(g-1) + 1:gR_G, 1:N_a]}$, s.t. $\mathbf{H}_{\mathrm{TX},l}^{(g)} \in \mathbb{C}^{R_G\times N_a}$, and $\mathbf{E}_{l,k}^{(g)}\in \mathbb{C}^{M\times N_a}$, where $g \in \{1,2, \dots, G\}$ denotes the group, and $R_G$ the group size.
The group-wise equivalent downlink channel matrix from all \acp{AP} to the $k$-th user is obtained following the method described in equation \eqref{eq:equivalent_channel}.
As such, to obtain a solution for the \ac{BD-RIS} scattering matrix design, the gradient of the objective function with respect to $\mathbf{\Theta}_g$ is required.

For convenience, the corresponding gradient is derived as a closed-form expression, given in \eqref{eq:grad_theta}, seen at the top of the next page. 
For clarity, the concatenated equivalent channel in \eqref{eq:equivalent_channel} can be rewritten as
\begin{equation}
\label{eq:eq_channel}
    \mathbf{E}_k = \mathbf{\bar{H}}_k + \mathbf{H}_{\mathrm{RX},k} \mathbf{\Theta} \mathbf{\bar{H}}_{\mathrm{TX}},
\end{equation}
where the \ac{LoS} channel matrix from all \acp{AP} to the $k$-th user is denoted as $\mathbf{\bar{H}}_k = [\mathbf{H}_{1,k}, \mathbf{H}_{2,k}, \dots, \mathbf{H}_{L,k}] \in \mathbb{C}^{M\times N_t}$, and the aggregated \ac{AP}-to-\ac{BD-RIS} channel is given by $\mathbf{\bar{H}}_{\mathrm{TX}} = [\mathbf{H}_{\mathrm{TX},1}, \mathbf{H}_{\mathrm{TX},2}, \dots, \mathbf{H}_{\mathrm{TX},L}] \in \mathbb{C}^{R\times N_t}$.

\begin{figure*}[t]
\begin{equation}
\label{eq:grad_theta}
{
\nabla_{\mathbf{\Theta}_g}\breve{{\eta}}
=
\sum_{k\in \mathcal{K}}
\Big[
\mathbf{H}_{\mathrm{RX},k}^{(g)H}\mathbf{Y}_k
\big(\mathbf{I}_M + \mathbf{Z}_k\big)
\mathbf{V}_k^H
\mathbf{\bar{H}}_{\mathrm{TX}}^{(g)H}
-
\mathbf{H}_{\mathrm{RX},k}^{(g)H}\mathbf{Y}_k
\big(\mathbf{I}_M + \mathbf{Z}_k\big)
\mathbf{Y}_k^H
\mathbf{E}_k
\Big(
\sum_{k'\in \mathcal{K}}\mathbf{V}_{k'}\mathbf{V}_{k'}^H
\Big)
\mathbf{\bar{H}}_{\mathrm{TX}}^{(g)H}
\Big]
}
\end{equation}
\begin{center}
\rule{1\textwidth}{0.1pt}
\end{center}
\end{figure*}}

{As a result of having obtained the scattering matrix}, the beamforming matrix $\mathbf{V}_k$ depends only on the effective channel $\mathbf{E}_k$.
Therefore, (P1) can be rewritten as
\begin{subequations}
    \begin{eqnarray}
    \text{(P2)}:  \quad  \underset{\mathbf{V}_{k}}{\mathrm{maximize}} && \sum_{k\in\mathcal{K}} {\breve{{\eta}}_k} \\ 
    \text{subject to} &&  \sum_{k\in\mathcal{K}}\|\mathbf{V}_k\|^2_\mathrm{F} \leq P_{\max} , \\
    && \|\mathbf{Q}_l\|^2_\mathrm{F} \leq P_{l}, \; \forall l \in \mathcal{L}.
    \end{eqnarray}
  \end{subequations}

Furthermore, the problem can be expressed solely in terms of $\mathbf{Q}_{l,k}$ sub-beamformers as
\begin{subequations}
\begin{eqnarray}
    \text{(P3)}:  \quad\underset{\mathbf{Q}_{l,k}}{\mathrm{maximize}} && \sum_{k\in\mathcal{K}} {\breve{{\eta}}_k} \\ 
    \text{subject to} && \sum_{k\in\mathcal{K}}\|\mathbf{Q}_{l,k}\|^2_\mathrm{F} \leq P_{l}, \; \forall l \in \mathcal{L}.
\end{eqnarray}
\end{subequations}

Based on the previous problem, the allocation of power to each AP based on the global available power and the power-beamforming direction breakdown, a \ac{FP} \ac{SCO} can be written as
\begin{subequations}
\begin{eqnarray}
\hspace{-2ex}
    \text{(P4)}:  \quad\underset{\mathbf{Q}_{l,k},\, p_{l,k}{,\, \nu_l}}{\mathrm{maximize}} && \sum_{k\in\mathcal{K}} {\breve{{\eta}}_k}\label{eq:p4_obj} \\
    \text{subject to} &&  \|\bar{\mathbf{Q}}_{l,k}\|^2_\mathrm{F} = 1, \; \forall l, k \in \mathcal{L}, \mathcal{K}, \label{eq:p4_q_lk}\\
    &&  \sum_{k\in\mathcal{K}} p_{l,k} \leq  P_{l}, \; \forall l \in \mathcal{L},
\end{eqnarray}
\end{subequations}
where 
\begin{equation}
\label{eq:q_lk}
    \mathbf{Q}_{l,k} = \sqrt{p_{l,k}} \bar{\mathbf{Q}}_{l,k}.
\end{equation}

Finally, $\breve{{\eta}}_k${, as seen in \eqref{eq:fulleta},} can be set as
\begin{equation}
    \breve{{\eta}}_k = 2 \cdot \mathrm{Tr}\Big( \Re \{  \mathbf{V}_{k}^\mathrm{H} \mathbf{A}_k\} \Big) - \mathrm{Tr}\Big( \sum\limits_{k'\in \mathcal{K}} \mathbf{V}_{k'} \mathbf{V}_{k'}^\mathrm{H} \mathbf{B}_{k} \Big),
\end{equation}
where for convenience, equivalent matrices $\mathbf{A}_k$ and $\mathbf{B}_k $ under the rotational property of the trace have been introduced as
\begin{eqnarray}
    && \mathbf{A}_k = \mathbf{E}^\mathrm{H}_{k} \mathbf{Y}_k \big( \mathbf{I}_{M}+\mathbf{Z}_k \big), \label{eq:A_k}\\
    && \mathbf{B}_k = \mathbf{E}^\mathrm{H}_{k} \mathbf{Y}_k \big( \mathbf{I}_{M}+\mathbf{Z}_k \big) \mathbf{Y}_k^\mathrm{H} \mathbf{E}_{k}. \label{eq:B_k}
\end{eqnarray}
Under the definition of \eqref{eq:vk_from_qlk}, the two trace terms can be rewritten as
\begin{align}
    \breve{{\eta}}_k & = 2 \sum\limits_{l\in \mathcal{L}} \! \mathrm{Tr} \Big( \Re \{  \mathbf{Q}_{l,k}^\mathrm{H} \big[\mathbf{A}_k\big]_l \} \Big)  \\
    &\hspace{2ex} -  \sum\limits_{k'\in \mathcal{K}}\sum\limits_{l\in \mathcal{L}}\sum\limits_{l'\in \mathcal{L}} \! \mathrm{Tr}\Big( \mathbf{Q}_{l,k'} \mathbf{Q}_{l',k'}^\mathrm{H} \big[\mathbf{B}_k\big]_{l',l} \Big), \nonumber
\end{align}
where $\big[ \cdot \big]_{l}$ selects the $l$-th row group consisting of $N_a$ rows, and $\big[ \cdot \big]_{l',l}$ selects the $l',l$-th submatrix of size $N_a \times N_a$.

As such, the sum rate can be expressed as
\begin{align}
    \sum\limits_{k' \in \mathcal{K}} \breve{{\eta}}_{k'} & = 2 \sum\limits_{k' \in \mathcal{K}} \sum\limits_{l'\in \mathcal{L}} \mathrm{Tr}\Big( \Re \{  \mathbf{Q}_{l',k'}^\mathrm{H} \big[\mathbf{A}_{k'} \big]_{l'} \} \Big) \\
    & \hspace{2ex} - \! \sum\limits_{k'\in \mathcal{K}} \sum\limits_{k'' \in \mathcal{K}}\sum\limits_{l'\in \mathcal{L}}\sum\limits_{l''\in \mathcal{L}}\! \mathrm{Tr}\Big( \mathbf{Q}_{l',k''} \mathbf{Q}_{l'',k''}^\mathrm{H} \big[\mathbf{B}_{k'}\big]_{l'',l'} \Big). \nonumber
\end{align}

The first step towards the closed-form \ac{FP} solution is to derive the sum rate with respect to $\mathbf{Q}_{l,k}$, as shown at the top of the next page of the article in \eqref{eq:rate_derivative}.

\begin{figure*}
\begin{equation}
\label{eq:rate_derivative}
    \frac{\partial}{\partial \mathbf{Q}_{l,k}}\sum\limits_{k'\in \mathcal{K}} \breve{{\eta}}_{k'} = 2 \big( \big[\mathbf{A}_{k} \big]_{l} \big)^\mathrm{H} \!\! - 2 \mathbf{Q}_{l,k}^\mathrm{H} \!\!\sum\limits_{k'\in \mathcal{K}}\!\! \big[\mathbf{B}_{k'}\big]_{l,l} - \sum\limits_{k' \in \mathcal{K}} \sum\limits_{l' \in \mathcal{L} \backslash \{l\} } \!\!\!\!\!\! \Big( \! \big( \big[\mathbf{B}_{k'}\big]_{l,l'} \mathbf{Q}_{l',k} \big)^\mathrm{H} \!\! + \! \mathbf{Q}_{l',k}^\mathrm{H} \big[\mathbf{B}_{k'}\big]_{l',l}  \Big)
\end{equation}
\begin{center}
    \rule{1\textwidth}{0.1pt}
\end{center}
\end{figure*}

Following the power constraints used in \cite{SandovalA2023}, the closed form can be obtained as
\begin{equation}
\label{eq:closed_form_q_lk}
    \mathbf{Q}_{l,k} = \Big( \zeta_{l} \mathbf{I}_{N_a} + \sum\limits_{k'\in \mathcal{K}} \big[\mathbf{B}_{k'}\big]_{l,l} \Big)^{-1} \Big( \big[\mathbf{A}_{k} \big]_{l} - \frac{1}{2} \mathbf{\Omega}_{l,k} \Big),
\end{equation}
with
\begin{equation}
\label{eq:omega_lk}
    \mathbf{\Omega}_{l,k} = \sum\limits_{k' \in \mathcal{K}} \sum\limits_{l' \in \mathcal{L} \backslash \{l\} } \Big( \big[\mathbf{B}_{k'}\big]_{l,l'} +  \big[\mathbf{B}_{k'}\big]_{l',l}^\mathrm{H}  \Big) \hspace{-4.5ex} \underbrace{\mathbf{Q}_{l',k}}_{\begin{subarray}{c}\text{\tiny Treated as constant, updated} \\ \text{\tiny with minorizing matrices} \end{subarray}}.
\end{equation}

\begin{figure}[H]
\centering
  \includegraphics[width=1\columnwidth]{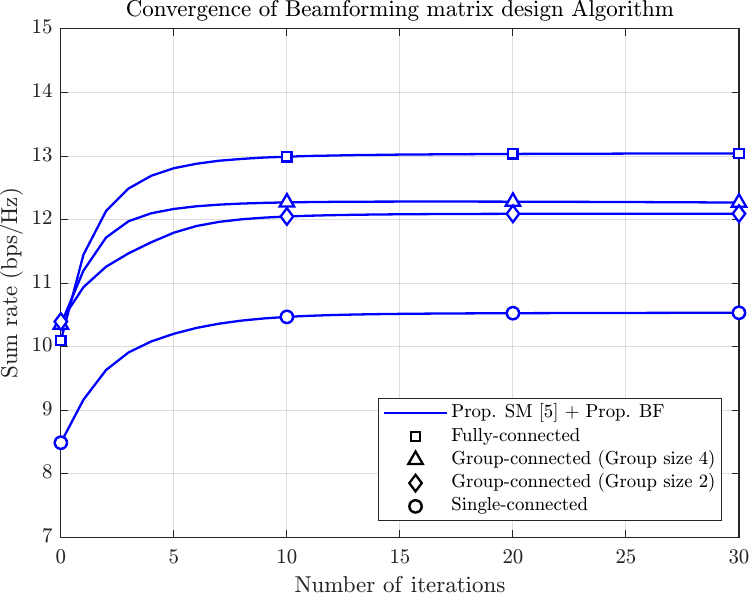}
  \caption{Convergence of Algorithm 1 vs. number of iterations for different connectivity structures{, where $R = 32$, $M = 2$, $K = 4$, $N_a = 2$, $L = 4$ and $P_{\mathrm{max}} = 16$\,dBm.}}
  \label{fig:convergence}
\vspace{-2ex}
\end{figure}

For the convenience of the reader, the full procedure implemented in this article is outlined in Algorithm \ref{alg:BF}, and a convergence plot is provided in Figure \ref{fig:convergence}, {where it can be observed that the proposed \ac{BF} algorithm converges rapidly, requiring $\sim10$ iterations for all considered \ac{BD-RIS} architectures, while higher-connectivity configurations achieve larger final sum-rate values without affecting the convergence speed.}

\begin{algorithm}[H]
\caption{{Proposed \ac{FP}-based per-\ac{AP} \ac{BF} Optimization}}
\label{alg:BF}
\begin{algorithmic}[1]
\REQUIRE $\mathbf{E}$, $P_{\mathrm{max}}$, $N_0$, $i_{\mathrm{max}}^{\mathrm{FP}}$, {$\zeta_{\mathrm{min}}, \zeta_{\mathrm{max}}$}, $\epsilon$
\ENSURE TX-beamforming matrix $\mathbf{V}$
\INITIALIZE MMSE-based $\mathbf{V}^{(0)}$, $\mathbf{Q}^{(0)}_{l,k} \ \forall l,k$ from $\mathbf{V}_k^{(0)} \ \forall k$ , $i^{\mathrm{FP}} \gets 0$
\REPEAT
    \STATE $i^{\mathrm{FP}} \gets i^{\mathrm{FP}} + 1$
    \STATE Compute $\mathbf{Z}_k$, $\forall k$ from \eqref{eq:z_k}
    \STATE Compute $\mathbf{Y}_k$, $\forall k$ from \eqref{eq:y_k}
    \STATE Compute $\mathbf{A}_k$, $\forall k$ from \eqref{eq:A_k}
    \STATE Compute $\mathbf{B}_k$, $\forall k$ from \eqref{eq:B_k}
    \STATE Obtain $\mathbf{\Omega}_{l,k}$, $\forall l,k$, following \eqref{eq:omega_lk}
    \STATE Obtain $\mathbf{Q}_{l,k}^{(i^{\mathrm{FP}})}$, $\forall l,k$, with optimized {$\zeta_{l}$} via bisection search over $[{\zeta_{min}}, {\zeta_{max}}]$ to satisfy \eqref{eq:p1_sbf}
    \STATE Compute $\mathbf{V}_k$, $\forall k$ from \eqref{eq:vk_from_qlk}
\UNTIL{$\bigl\| \mathbf{V}^{(i^{\mathrm{FP}})} - \mathbf{V}^{(i^{\mathrm{FP}} - 1)} \bigr\|_{\mathrm{F}} < \epsilon,\ \forall k \ \textbf{or} \ i^{\mathrm{FP}} = i^{\mathrm{FP}}_{\max}
$}
\RETURN$\mathbf{V}$
\end{algorithmic}
\end{algorithm}

\textit{Computational Complexity:} { For the sake of comparison against existing \ac{SotA} methods, a full computational complexity review is performed.
The computational complexity of Algorithm \ref{alg:BF} is mainly influenced by the number of bisection-search and \ac{FP} iterations, i.e., $i^{\mathrm{BS}}$ and $i^{\mathrm{FP}}$, in combination with computation of the sub-\ac{BF} matrices.
As such, for the computation of $\mathbf{Q}_{l,k}\ \forall l,k$, two loops over the $L$ \acp{AP} and $K$ users are required, yielding a complexity of $\mathcal{O}\big(KL\big)$.
Furthermore, in the worst case scenario, the inverse of the $N_a\times N_a$ matrix, yields a complexity of $\mathcal{O}\big(N_a^3\big)$.
The inverse is then followed by an $N_a\times M$ matrix multiplication, which yields a  total complexity for the computation of all sub-\ac{BF} matrices equal to 
\begin{equation}
    \mathcal{O}\big(i^{FP}i^{BS}KLN_a^2\big(N_a + M\big)\big).
\end{equation}

Another source of considerable impact would be the computation of $\mathbf{\Omega}_{l,k}$, where four loops are considered to compute the full $\mathbf{\Omega}$ matrix, namely two loops going to $K$, one going to $L$ and another going to $L-1$, resulting in a complexity of $\mathcal{O}\big(K^2L^2\big)$.
The computation of $\mathbf{\Omega}$, further includes a multiplication of two matrices of size $N_a\times N_a$ and $N_a\times M$, yielding a total complexity of $\mathcal{O}\big(K^2L^2N_a^2M\big)$.
Taking account of all of the contributions, yields a total computational complexity of Algorithm \ref{alg:BF} of
\begin{equation}
    \mathcal{O}\big(i^{FP}\big(i^{BS}KLN_a^2\big(N_a +  M\big) + K^2L^2N_a^2M\big)\big).
    \label{eq:complexity}
\end{equation}

}


\vspace{-3ex}

\section{Simulation Results}
\label{sec:simresults}

{This section evaluates the effectiveness of the proposed \ac{BF} matrix design in improving the communication performance of \ac{RBD-RIS}-aided \ac{CF}-\ac{MImMO} systems through computer simulations conducted under the following parameters.\footnote{{In all figures, except Figure \ref{fig:MUMISO}, the scattering matrix is designed only following the approach in \cite{FidanovskiTWC2026}.
Since this work extends the aforementioned method, a direct extension of the comparisons presented therein is also provided.}}

The number of users and \acp{AP} depends on the considered scenario. Specifically, the performance of fully-, over-, and under-loaded systems is examined.
The \ac{RIS} structure consists of $R = 32$ \acp{RE}, unless otherwise stated, and the parameters used for the scattering matrix design are identical to those in \cite{FidanovskiTWC2026}; furthermore, each \ac{AP} is subject to a maximum transmit power $P_{\mathrm{AP}} = P_{\mathrm{max}}/L$, such that the total transmit power available to the system, $P_{\mathrm{max}}$, ranges from $0$ to $16\,\mathrm{dBm}$.
{The large-scale fading coefficients $\Upsilon(d)$ are generated using the \ac{3GPP} \ac{UMi} path-loss models described in \cite[Table B.1.2.1-1]{3gpp36814}. 
Specifically, the \ac{AP}-to-\ac{BD-RIS} and \ac{BD-RIS}-to-user links are modeled as favorable \ac{LoS} links, while the direct \ac{AP}-to-user link is modeled as a strongly attenuated \ac{NLoS} link\footnote{This configuration highlights the role of the \ac{BD-RIS} in enhancing the propagation environment and is thus of practical relevance.}.
The small-scale fading is modeled as Rician fading for the \ac{LoS} \ac{AP}-to-\ac{BD-RIS} and \ac{BD-RIS}-to-user links, with Rician factor $K_f = 9\,\mathrm{dB}$, while the direct \ac{AP}-to-user link is modeled as Rayleigh fading, corresponding to $K_f = 0$, in accordance with \cite[Table B.1.2.2.1-4]{3gpp36814}.

}

The link distances are set as $d = 2.5\mathrm{m}$ for the \ac{BD-RIS}-to-user link, $d = 50\,\mathrm{m}$ for the \ac{AP}-to-\ac{BD-RIS} link, and $d = 51\,\mathrm{m}$ for the \ac{AP}-to-user link. 
The system operates at a carrier frequency of $2.4\mathrm{GHz}$, and each user experiences identical noise with power spectral density $N_0 = -80\,\mathrm{dBm}$.

The first set of results is provided in Figure \ref{fig:CDF_sR}, and aims to demonstrate the effectiveness of the proposed \ac{BF} matrix design against \ac{SotA} methods \cite{SandovalA2023, MirettiTSP2024}.
The scattering matrix ``SM'' follows the design proposed in \cite{FidanovskiTWC2026}, while the \ac{SotA} \ac{BF} schemes considered are the \ac{FP} \ac{BF} matrix from \cite{SandovalA2023}, and the \ac{MMSE} \ac{BF} with per-\ac{AP} power normalization \cite{MirettiTSP2024}. 
The proposed \ac{BF} design demonstrates clear performance improvements compared to both \ac{SotA} methods at a transmit power of $P_{\mathrm{max}} = 16\,\mathrm{dBm}$. 
In particular, the combination of the proposed \ac{BF} with the single-connected ``SC'' scattering matrix achieves notable sum-rate gains.

The second set of results, namely Figure \ref{fig:CDF_Uk_R}, complements Figure \ref{fig:CDF_sR} by illustrating the per-user rate performance of the proposed \ac{BF} matrix design in combination with the scattering matrix ``SM'' design from \cite{FidanovskiTWC2026}, considering the {single-connected architecture}.
The proposed method achieves superior per-user rates over most of the distribution, with only marginal degradation in a limited region. 
{These gains mainly result from enforcing the per-\ac{AP} power constraints within the sum-rate maximization.}
Despite this, the average performance remains consistently higher than that of the considered \ac{SotA} methods.

\begin{figure}[H]
    \centering
    \includegraphics[width=1\columnwidth]{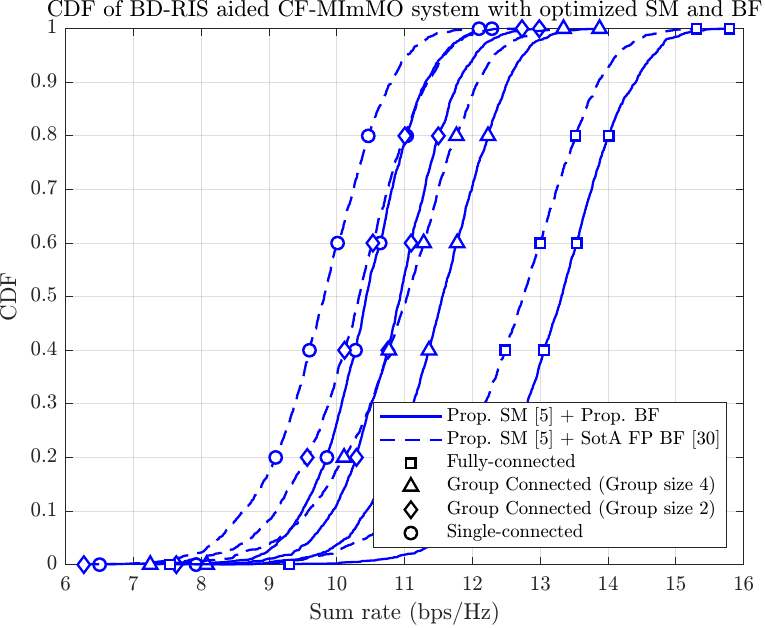}
    \caption{CDF of sum-rate performance of the proposed vs. SotA \ac{BF} matrix design \cite{SandovalA2023} with the fully-connected ``FC'', group-connected ``GC'', with group size of 2 and 4, and the single-connected ``SC'' architecture{, where $R = 32$, $M = 2$, $K = 4$, $N_a = 2$, $L =  4$, $\nu = 1$, and $P_{\mathrm{max}} = 16$\,dBm.}}
    \label{fig:CDF_sR}
    \vspace{-3ex}
\end{figure}

\begin{figure}[H]
    \centering
    \includegraphics[width=1\columnwidth]{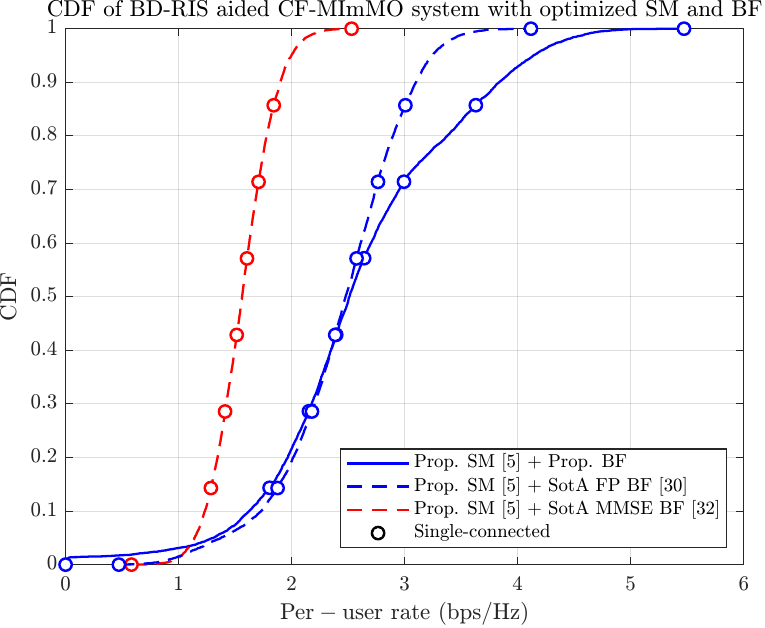}
    \caption{{CDF of per-user rate performance of the proposed vs. SotA \ac{BF} matrix designs \cite{SandovalA2023, MirettiTSP2024}, considering {the single-connected ``SC'' architecture, where $R = 32$, $M = 2$, $K = 4$, $N_a = 2$, $L =  4$, $\nu = 1$, and $P_{\mathrm{max}} = 16$\,dBm.}}}
    \label{fig:CDF_Uk_R}
    \vspace{-2ex}
\end{figure}

\vspace{-0.5ex}

Further assessments aim to demonstrate the performance of the proposed \ac{BF} matrix design in combination with the proposed scattering matrix ``SM'' over a transmit power $P_{\mathrm{max}}$ range, i.e., $0-16$\,dBm,  under different loaded cases, namely over-, fully-, and under-loaded scenarios, respectively.

Namely, in Figure \ref{fig:SCvsFC} the sum rate performance of the proposed \ac{BF} matrix design in combination with the single-connected ``SC'' and fully-connected ``FC'' scattering matrix \cite{FidanovskiTWC2026} is showcased.
Complementary, Figure \ref{fig:GC} assesses the sum-rate performance of the proposed \ac{BF} matrix design in combination with the group-connected scattering matrix, with group size 2 and 4, ``GC(2)'' and ``GC(4)'', respectively.

\begin{figure}[H]
  \centering
  \subfloat[Over-loaded {($K = 6$)}\label{fig:SC_over}]{
    \includegraphics[width=\columnwidth]{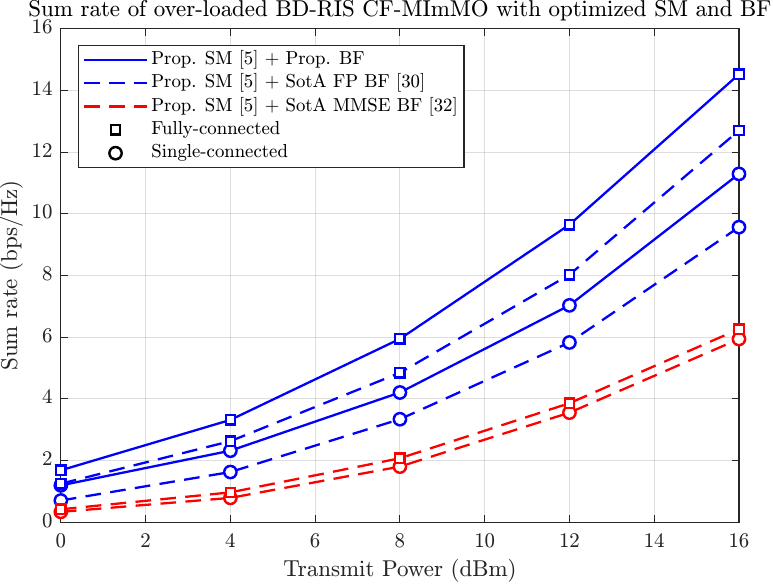}
  }\\ \vspace{-0.1ex}
  \subfloat[Fully-loaded {($K = 4$)}\label{fig:SC_fully}]{
    \includegraphics[width=\columnwidth]{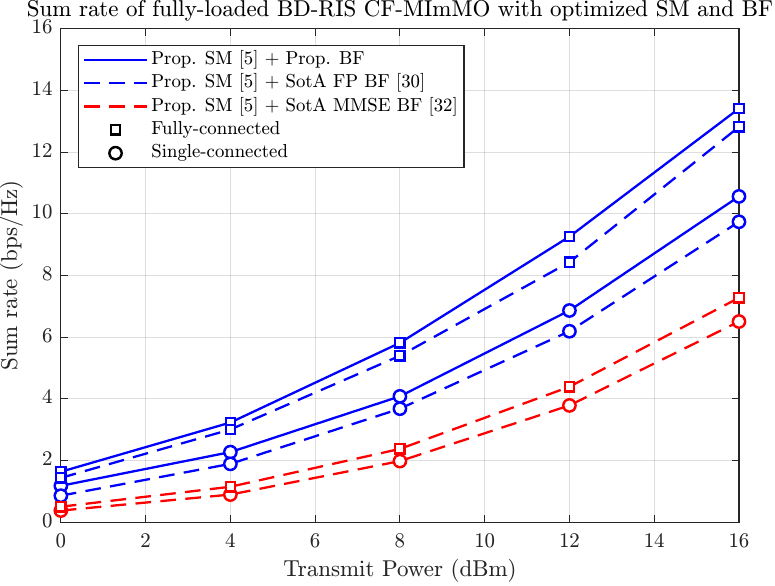}
  }\\ \vspace{-0.1ex}
  \subfloat[Under-loaded {($K = 3$)}\label{fig:SC_under}]{
    \includegraphics[width=\columnwidth]{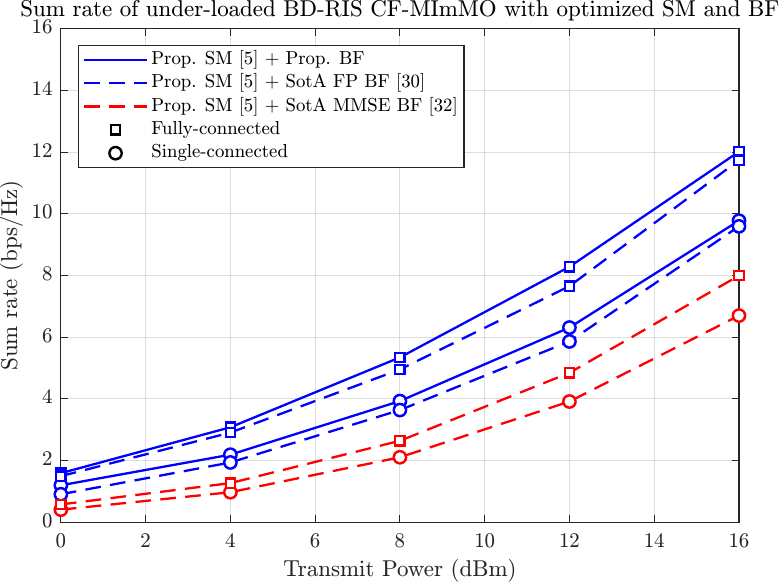}
  }
  \caption{Comparison of the sum-rate performance between the proposed method and the \ac{SotA} \ac{BF} matrix designs \cite{SandovalA2023, MirettiTSP2024} with the fully-, and single-connected, ``FC'' and ``SC'' architecture, under over-, fully-, and under-loaded system scenarios{, where $R = 32$, $M = 2$, $N_a = 2$, $L = 4$, $\nu = 1$.}}
  \label{fig:SCvsFC}
\end{figure}

\begin{figure}[H]
  \centering
  \subfloat[Over-loaded {($K = 6$)}\label{fig:GC_over}]{
    \includegraphics[width=\columnwidth]{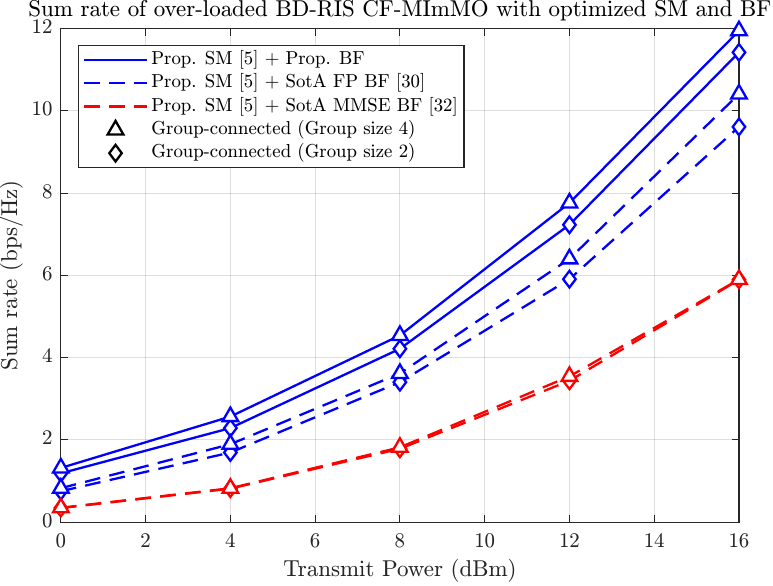}
  }\\ \vspace{-0.1ex}
  \subfloat[Fully-loaded {($K = 4$)}\label{fig:GC_fully}]{
    \includegraphics[width=\columnwidth]{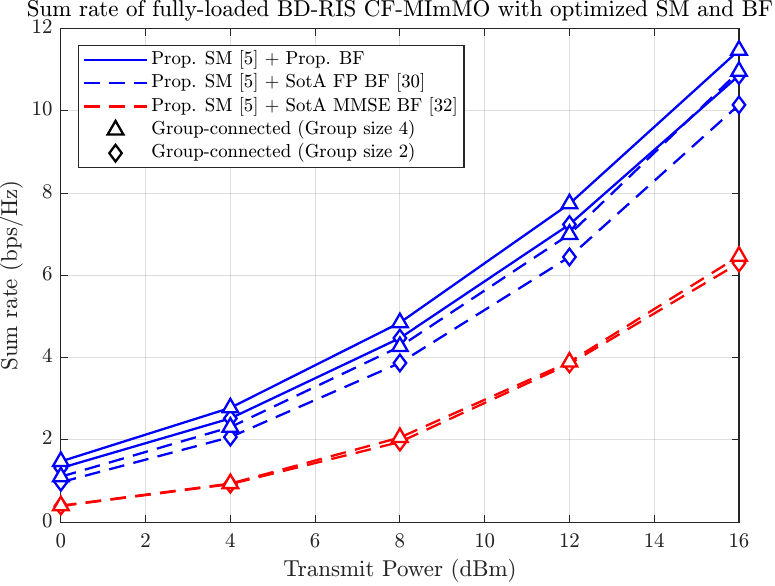}
  }\\ \vspace{-0.1ex}
  \subfloat[Under-loaded {($K = 3$)}\label{fig:GC_under}]{
    \includegraphics[width=\columnwidth]{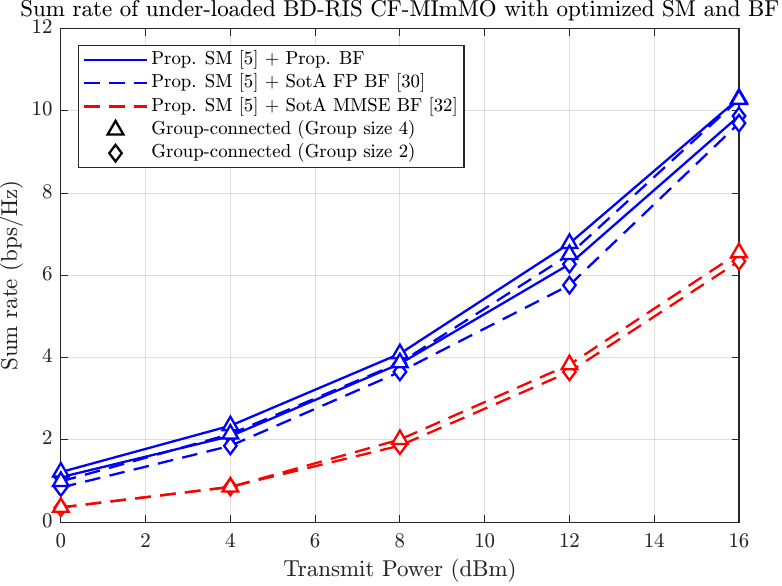}
  }
  \caption{Comparison of the sum-rate performance between the proposed method and the \ac{SotA} \ac{BF} matrix designs \cite{SandovalA2023, MirettiTSP2024} with the group-connected ``GC'' architecture, with a group size 2 and 4, under over-, fully-, and under-loaded system scenarios{, \!\! where $R = 32$, $M = 2$, $N_a = 2$, $L = 4$, $\nu = 1$.}}
  \label{fig:GC}
\end{figure}

The proposed \ac{BF} matrix design exhibits consistent performance gains across different system loading regimes, including over-, fully-, and under-loaded scenarios, compared to the considered \ac{SotA} methods.
{The most pronounced gains, together with the highest overall sum-rate values, are observed in the overloaded regime.
In this case, the larger number of served users makes the system more interference-limited, which increases the importance of coordinating the distributed transmit resources under the per-\ac{AP} power constraints.
The proposed design performs this coordination more effectively than the considered \ac{SotA} methods, leading to improved inter-user interference mitigation and higher performance.
Nevertheless, the performance differences among the under-, fully-, and over-loaded cases remain relatively small.
This indicates that, when many users are present, comparable average performance can be achieved by serving only a subset of users at a time, for example, through round-robin scheduling.
Such an approach can reduce the system complexity while maintaining nearly the same performance level, since the overall sum-rate is not strongly affected by the considered loading regime.}


{As previously mentioned, the work done in this article constitutes and extension of \cite{FidanovskiTWC2026}, through the design of the \ac{BF} matrix for the proposed scattering matrix design.
{For consistency, the same channel model as in \cite{FidanovskiTWC2026} is adopted in the subsequent numerical evaluations, where the direct \ac{AP}-to-user link is omitted.}
Therefore, the following figures present results that directly extend the comparisons reported in \cite{FidanovskiTWC2026} by incorporating the proposed \ac{BF} matrix design into the comparisons presented in that article.


As such, Figure \ref{fig:MUMISO} presents a comparison of the sum-rate performance of the proposed scattering matrix ``SM'' \cite{FidanovskiTWC2026}, in combination with either uniform power allocation ``PA'', or the proposed \ac{BF}, against the joint scattering and beamforming matrix design based on the \ac{pp-ADMM} framework\footnote{{It should be noted that the aforementioned framework was originally developed for the \ac{MU}-\ac{MISO} scenario and therefore requires additional modifications to be applicable in the \ac{CF}-\ac{MIMO} setting.
In contrast, the proposed framework can directly accommodate the less distributed \ac{MU}-\ac{MISO} case through appropriate selection of the system parameters.}} \cite{WuArx2024}.
While the proposed \ac{BF} design combined with the single-connected ``SC'' architecture exhibits a performance loss relative to the \ac{pp-ADMM} method, the additional architectures considered, namely the group-connected ``GC'' architecture with group size 4 and the fully-connected ``FC'' architecture, achieve similar or slightly better performance, with similar computational complexity.\footnote{{As shown in \cite{FidanovskiTWC2026}, the scattering matrix ``SM'' design, when combined with \ac{SotA} \ac{BF} methods, has a computational complexity comparable to that of the \ac{pp-ADMM} framework.
In the present work, however, we build upon the framework developed in \cite{FidanovskiTWC2026}, which introduces an additional additive term in the complexity, as given in \eqref{eq:complexity}. Under the sparse system setting considered in this comparison, this term is negligible relative to the computational complexity of the SM design in \cite{FidanovskiTWC2026}.}}
Moreover, the considered system model, i.e., the \ac{CF}-\ac{MImMO} setting, together with the more general \ac{BF} design, which is not restricted to the \ac{MU}-\ac{MISO} case, provides greater flexibility and broader applicability of the proposed framework. 
In addition, the reported results include scenarios with a larger number of \ac{RX} antennas, $M$, thereby illustrating the behavior of the proposed approach in a more general \ac{MIMO} setting.
}

\newpage

\begin{figure}[H]
  \centering
  \subfloat[Single-connected\label{fig:SC_MUMISO}]{
    \includegraphics[width=\columnwidth]{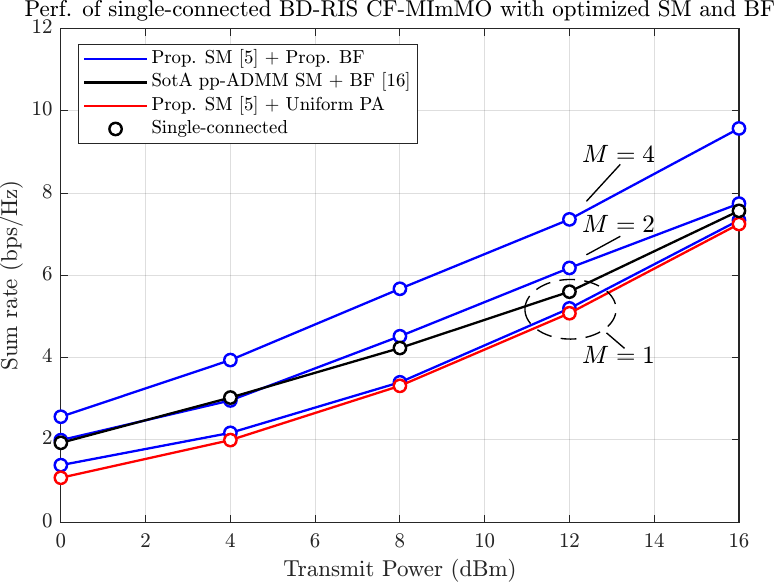}
  }\\ \vspace{-0.1ex}
  \subfloat[Group-connected\label{fig:GC_MUMISO}]{
    \includegraphics[width=\columnwidth]{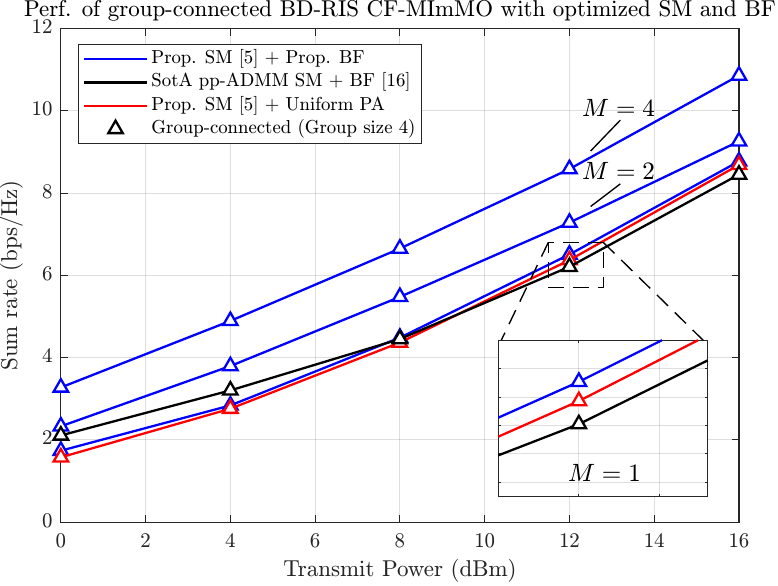}
  }\\ \vspace{-0.1ex}
  \subfloat[Fully-connected\label{fig:FC_MUMISO}]{
    \includegraphics[width=\columnwidth]{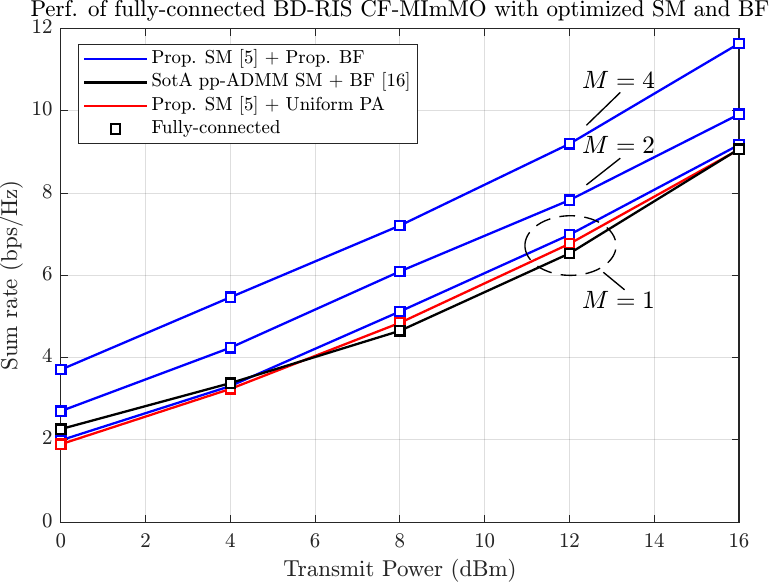}
  }
  \caption{{Comparison of the sum-rate performance using joint scattering matrix ``SM'' and ``\ac{BF}'' schemes, considering the proposed and \ac{SotA} \cite{WuArx2024} fully-connected ``FC'', group-connected ``GC'' architecture, with a group size 4, and single-connected ``SC'' architecture{, where $R = 32$, $M \in \{1,2,4\}$, $K = 2$, $N_a = 2$, $L = 1$, and $\nu = 1$.}}}
  \label{fig:MUMISO}
\end{figure}

\begin{figure}[H]
    \centering
    \includegraphics[width=1\columnwidth]{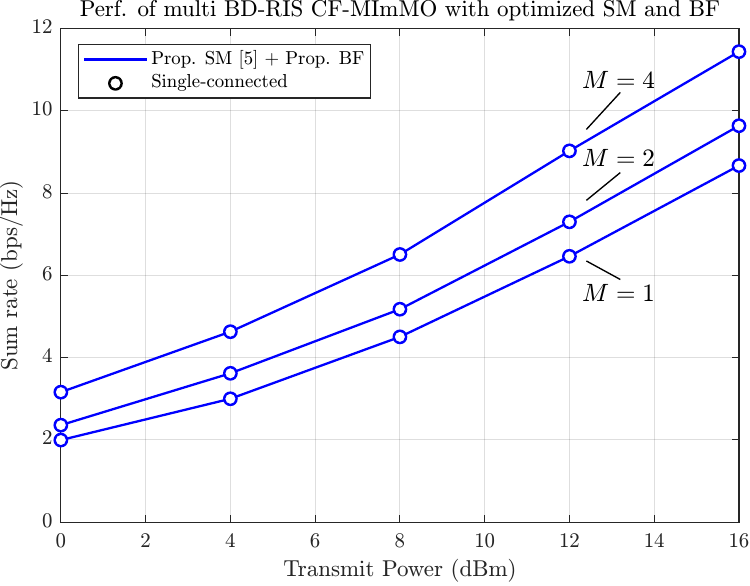}
    \caption{{Sum-rate performance of the proposed \ac{BF} matrix design in a setting with $B$ deployed \acp{BD-RIS}, considering the single-connected ``SC'' architecture{, where $R = 16$, $B = 2$, $M \in  \{1,2,4\}$, $K = 2$, $N_a = 2$, $L =  1$, and $\nu = 1$.}}}
    \label{fig:multi_RIS}
\end{figure}

\vspace{-1ex}

{Lastly, in Figure \ref{fig:multi_RIS}, we present the sum-rate performance of the proposed \ac{BF} matrix design in a scenario with multiple, $B$, \acp{BD-RIS}.\footnote{{It should be noted that, although \cite{KatsanosArx2026} considers a related multi-\ac{BD-RIS} \ac{CF}-\ac{mMIMO} setting, the system model and optimization framework differ materially from those considered herein. 
In particular, \cite{KatsanosArx2026} studies a decentralized wideband \ac{OFDM} design in which \ac{CSI} is locally acquired and exchanged among the \acp{BS}, while the final \ac{BD-RIS} configurations are obtained through consensus-based updates. 
By contrast, our work adopts a centralized design framework and focuses on \ac{FP}-based per-\ac{AP} beamforming together with direct scattering matrix optimization. 
Therefore, including \cite{KatsanosArx2026} as a numerical benchmark would not constitute a direct and fair comparison.}} 
The simulation parameters for the multi-\ac{BD-RIS} case are selected such that the deployed surfaces remain of comparable large-scale strength, either by retaining the baseline distances of the single \ac{BD-RIS} setup or by introducing only mild asymmetry. 
This choice is made to highlight the gains enabled by the additional spatial flexibility of multiple deployed \acp{BD-RIS}, rather than a regime in which one surface becomes effectively dominant.
The results show that the proposed design can effectively exploit the additional \acp{DoF} offered by multiple deployed \acp{BD-RIS} to further improve system performance.
The results show that the proposed design can effectively exploit the additional \acp{DoF} offered by multiple deployed \acp{BD-RIS} to further improve system performance.
More specifically, for the same system parameters as in Figure \ref{fig:MUMISO}, but with one additional \ac{BD-RIS} and under the single-connected ``SC'' architecture, the achieved sum-rate becomes comparable to that attained by the fully-connected ``FC'' architecture. 
In particular, the combination of two smaller single-connected \acp{BD-RIS}, each comprising $16$ \acp{RE}, is able to deliver similar performance to that of a single fully-connected \ac{BD-RIS} with $32$ \acp{RE}. 
This highlights that distributing the available \acp{RE} across multiple simpler surfaces can provide substantial performance gains, while avoiding the increased architectural complexity associated with a single larger fully-connected \ac{BD-RIS}.

It is further noted that the proposed framework is sufficiently general to accommodate heterogeneous multi-\ac{BD-RIS} deployments, such that different surfaces may adopt different architectures, including single-connected, group-connected, and fully-connected configurations.
Nevertheless, in this work, two single-connected \acp{BD-RIS} are considered to facilitate a direct comparison with the previously presented results.
}

\section{Conclusion}
\label{sec:conc}
{A novel sum-rate maximization \ac{BF} matrix design scheme for \ac{RBD-RIS}-aided \ac{CF}-\ac{MIMO} systems has been presented in this article.
The proposed design is based on the \ac{FP} framework and is developed to satisfy per-\ac{AP} power constraints, while remaining applicable to a wide range of scattering matrix designs.
In addition, the scattering matrix design method proposed in prior work has been extended to the more general \ac{MIMO} setting, and a closed-form gradient expression has been derived, enabling efficient optimization via Riemannian manifold-based methods.
Simulation results demonstrate the effectiveness of the proposed approach in improving the communication performance of \ac{RBD-RIS}-aided \ac{CF}-\ac{MIMO} systems compared to \ac{SotA} methods, across different system loading regimes, including over-, fully-, and under-loaded scenarios.

The results further highlight that the proposed design achieves consistent performance gains at both the system and user levels, while effectively handling the distributed nature of the network.
In particular, the observed improvements in interference-limited regimes underline the importance of jointly optimizing the beamforming and scattering matrix under practical power constraints.
Moreover, the results suggest that exploiting the additional spatial flexibility offered by multiple less complex \acp{BD-RIS} may be more effective than increasing the architectural complexity of a single large \ac{BD-RIS}.
Overall, these findings confirm that the proposed framework provides a flexible and efficient solution for enhancing the performance of \ac{RBD-RIS}-aided systems.
}
}

\appendices

\section{Extension to multiple \acp{BD-RIS}}
\label{sec:appendix}

{

The system model in Section~\ref{sec:sysmodel} can be extended to the case of multiple deployed \acp{BD-RIS} by introducing a set of surfaces indexed by $b \in \{1,2,\dots,B\}$, where each $b$-th \ac{BD-RIS} is characterized by its scattering matrix $\mathbf{\Theta}^{(b)} \in \mathbb{C}^{R_b \times R_b}$, the channel $\mathbf{H}_{\mathrm{TX},l}^{(b)} \in \mathbb{C}^{R_b \times N_a}$ from the $l$-th \ac{AP} to the $b$-th \ac{BD-RIS}, and the channel $\mathbf{H}_{\mathrm{RX},k}^{(b)} \in \mathbb{C}^{M \times R_b}$ from the $b$-th \ac{BD-RIS} to the $k$-th user.
Then, the equivalent channel between the $l$-th \ac{AP} and the $k$-th user, which in the single-\ac{BD-RIS} case is given by $\mathbf{E}_{l,k}=\mathbf{H}_{l,k}+\mathbf{H}_{\mathrm{RX},k}\mathbf{\Theta} \mathbf{H}_{\mathrm{TX},l}$, is generalized as
\begin{equation}
\mathbf{E}_{l,k}=\mathbf{H}_{l,k}+\sum_{b=1}^{B} \mathbf{H}_{\mathrm{RX},k}^{(b)} \mathbf{\Theta}^{(b)} \mathbf{H}_{\mathrm{TX},l}^{(b)}.
\end{equation}

Accordingly, the aggregate downlink channel from all \acp{AP} to user $k$ becomes $\mathbf{E}_k=[\mathbf{E}_{1,k},\mathbf{E}_{2,k},\dots,\mathbf{E}_{L,k}]$, while the received signal model, {the interference-plus-noise-whitened signal covariance matrix}, and the subsequent \ac{FP}-based beamforming design remain unchanged in form after replacing the single-surface equivalent channel by the above multi-\ac{BD-RIS} expression. 
In this case, the optimization variables are extended from $\mathbf{\Theta}$ to the set $\{\mathbf{\Theta}^{(b)}\}_{b=1}^{B}$, with each scattering matrix satisfying its corresponding reciprocity, losslessness, and connectivity constraints. 
This extension therefore preserves the overall structure of the proposed framework, while allowing the joint modeling of multiple \ac{BD-RIS}-assisted propagation paths.

\noindent\textit{Approximate decoupling of the multi-BD-RIS contributions in the MIMO case:}
Let the effective channel of user $k$ be extended from the single-\ac{BD-RIS} model in \eqref{eq:eq_channel} to
\begin{equation}
\mathbf{E}_k = \bar{\mathbf{H}}_k + \sum_{b=1}^{B} \mathbf{H}_{\mathrm{RX},k}^{(b)} \mathbf{\Theta}^{(b)} \bar{\mathbf{H}}_{\mathrm{TX}}^{(b)},
\label{eq:multi_bdris_Ek}
\end{equation}
where, for convenience, we define
\begin{equation}
\mathbf{E}_k^{(0)} \triangleq \bar{\mathbf{H}}_k, \quad \mathbf{E}_k^{(b)} \triangleq \mathbf{H}_{\mathrm{RX},k}^{(b)} \mathbf{\Theta}^{(b)}
\bar{\mathbf{H}}_{\mathrm{TX}}^{(b)}, \quad b=1,\dots,B,
\label{eq:local_Ek_b}
\end{equation}
such that
\begin{equation}
\mathbf{E}_k=\sum_{b=0}^{B}\mathbf{E}_k^{(b)}.
\label{eq:Ek_sum_b}
\end{equation}

The corresponding rate at the $k$-th user is
\begin{equation}
\eta_k = \log_2\left|\mathbf{I}_M+\mathbf{\Gamma}_k\right|\!,
\text{ with } \, \mathbf{\Gamma}_k = \mathbf{V}_k^H \mathbf{E}_k^H \mathbf{\Psi}_k^{-1} \mathbf{E}_k \mathbf{V}_k
\label{eq:rate_gamma_multi}
\end{equation}
and
\begin{equation}
\mathbf{\Psi}_k = \sum_{i\neq k} \mathbf{E}_k\mathbf{V}_i\mathbf{V}_i^H\mathbf{E}_k^H + N_0\mathbf{I}_M.
\label{eq:Psi_multi}
\end{equation}

Since the matrix \ac{FP} reformulation expresses the objective through the matrices \eqref{eq:A_k} and \eqref{eq:B_k},
it is sufficient to study how these quantities depend on the individual \ac{BD-RIS} contributions.

From \eqref{eq:Ek_sum_b}, the matrix $\mathbf{A}_k$ admits the exact decomposition
\begin{equation}
\mathbf{A}_k = \sum_{b=0}^{B} \mathbf{A}_k^{(b)}, \quad
\mathbf{A}_k^{(b)} \triangleq \mathbf{E}_k^{(b)H}\mathbf{Y}_k(\mathbf{I}_M+\mathbf{Z}_k).
\label{eq:Ak_decomp_b}
\end{equation}

Similarly, $\mathbf{B}_k$ can be expanded exactly as
\begin{equation}
\mathbf{B}_k \!= \!\!\sum_{b=0}^{B}\!\sum_{c=0}^{B} \!\mathbf{B}_k^{(b,c)}\!\!, \,\, \mathbf{B}_k^{(b,c)} \!\!\triangleq \!\mathbf{E}_k^{(b)H}\mathbf{Y}_k(\mathbf{I}_M\!+\mathbf{Z}_k)\mathbf{Y}_k^H\mathbf{E}_k^{(c)}\!\!.
\label{eq:Bk_decomp_bc}
\vspace{-0.5ex}
\end{equation}

Hence, all coupling among different \acp{BD-RIS} appears only through the cross terms
\begin{equation}
\mathbf{B}_k^{(b,c)}, \qquad b\neq c,\;\; b,c\in\{1,\dots,B\}.
\label{eq:cross_terms_bc}
\vspace{-0.5ex}
\end{equation}

Assume now that the reflected contributions generated by different \acp{BD-RIS} are weakly coupled in the sense that
\begin{equation}
\left\|
\mathbf{B}_k^{(b,c)} \right\|_F \ll \left\| \mathbf{B}_k^{(b,b)} \right\|_F, \quad b\neq c,\;\; b,c\in\{1,\dots,B\},
\label{eq:weak_coupling_assumption}
\vspace{-0.5ex}
\end{equation}
for all $k$, which is reasonable when the deployed \acp{BD-RIS} are sufficiently separated in space, induce sufficiently distinct angular signatures, or are optimized so that their reflected fields are weakly correlated at the users. 

Finally, \eqref{eq:Bk_decomp_bc} can be rewritten as
\begin{align}
\mathbf{B}_k &=  \mathbf{B}_k^{(0,0)} +\sum_{b=1}^{B}\mathbf{B}_k^{(0,b)} \nonumber \\
& \,\,\,\,\,\,+\sum_{b=1}^{B}\mathbf{B}_k^{(b,0)} +\sum_{b=1}^{B}\mathbf{B}_k^{(b,b)} + \sum_{b=1}^{B}\sum_{\substack{c=1\\c\neq b}}^{B}\mathbf{B}_k^{(b,c)},
\label{eq:Bk_full_expand_split}
\end{align}
where under \eqref{eq:weak_coupling_assumption}, the matrix $\mathbf{B}_k$ is approximated as
\vspace{-1ex}
\begin{equation}
\mathbf{B}_k \approx \mathbf{B}_k^{(0,0)} + \sum_{b=1}^{B} \Big( \mathbf{B}_k^{(0,b)} + \mathbf{B}_k^{(b,0)} + \mathbf{B}_k^{(b,b)}
\Big),
\label{eq:Bk_local_approx}
\end{equation}
that is, the direct-path terms and the terms involving only the $b$-th \ac{BD-RIS} are retained, whereas the terms coupling different \acp{BD-RIS} are neglected.

Substituting \eqref{eq:Ak_decomp_b} and \eqref{eq:Bk_local_approx} into the reformualted objective
\begin{equation}
\breve{\eta}_k = 2\operatorname{Tr}\!\Big( \Re\big\{\mathbf{V}_k^H\mathbf{A}_k\big\} \Big) - \operatorname{Tr}\!\Big(
\sum_{i\in\mathcal{K}} \mathbf{V}_i\mathbf{V}_i^H\mathbf{B}_k
\Big),
\label{eq:fp_obj_user_k}
\end{equation}
yields
\begin{equation}
\breve{\eta}_k \approx \breve{\eta}_k^{(0)} + \sum_{b=1}^{B} \breve{\eta}_k^{(b)},
\label{eq:eta_local_sum}
\end{equation}
where $\breve{\eta}_k^{(0)}$ is independent of $\{\mathbf{\Theta}^{(b)}\}_{b=1}^{B}$ and
\begin{align}
\breve{\eta}_k^{(b)} &\triangleq 2\operatorname{Tr}\!\Big(
\Re\big\{\mathbf{V}_k^H\mathbf{A}_k^{(b)}\big\} \Big) \nonumber\\
&\quad- \operatorname{Tr}\!\Big( \sum_{i\in\mathcal{K}} \mathbf{V}_i\mathbf{V}_i^H \Big( \mathbf{B}_k^{(0,b)} + \mathbf{B}_k^{(b,0)} +
\mathbf{B}_k^{(b,b)} \Big) \Big).
\label{eq:eta_local_b}
\end{align}

Therefore, the approximate contribution of the $b$-th \ac{BD-RIS} depends only on
\begin{equation}
\mathbf{E}_k^{(b)} \triangleq \mathbf{H}_{\mathrm{RX},k}^{(b)} \mathbf{\Theta}^{(b)}
\bar{\mathbf{H}}_{\mathrm{TX}}^{(b)},
\label{eq:local_surface_channel_again}
\end{equation}
and on the direct channel $\bar{\mathbf{H}}_k$, but not on the channels of any other \ac{BD-RIS}.

\begin{figure*}[t]
\begin{equation}
{
\nabla_{\mathbf{\Theta}^{(b)}}\breve{\eta} \approx \sum_{k\in\mathcal{K}}
\Big[ \mathbf{H}_{\mathrm{RX},k}^{(b)H} \mathbf{Y}_k(\mathbf{I}_M+\mathbf{Z}_k) \mathbf{V}_k^H \bar{\mathbf{H}}_{\mathrm{TX}}^{(b)H} - \mathbf{H}_{\mathrm{RX},k}^{(b)H} \mathbf{Y}_k(\mathbf{I}_M+\mathbf{Z}_k) \mathbf{Y}_k^H \big( \bar{\mathbf{H}}_k + \mathbf{H}_{\mathrm{RX},k}^{(b)} \mathbf{\Theta}^{(b)} \bar{\mathbf{H}}_{\mathrm{TX}}^{(b)} \big) \Big( \sum_{i\in\mathcal{K}} \mathbf{V}_i\mathbf{V}_i^H \Big) \bar{\mathbf{H}}_{\mathrm{TX}}^{(b)H} \Big] }
\label{eq:local_gradient_b}
\end{equation}
\begin{center}
\rule{1\textwidth}{0.1pt}
\end{center}
\end{figure*}

Consequently, the gradient of the approximate objective with respect to $\mathbf{\Theta}^{(b)}$ depends only on the local \ac{CSI} of the $b$-th \ac{BD-RIS}. 
In particular, by keeping only the terms in \eqref{eq:eta_local_b} that depend on $\mathbf{\Theta}^{(b)}$, the gradient is obtained, as shown at the top of {this} page of the article, as seen in \eqref{eq:local_gradient_b}.

It has been shown that, after neglecting the inter-\ac{BD-RIS} coupling terms $\mathbf{B}_k^{(b,c)}$ for $b\neq c$, the passive update of the $b$-th \ac{BD-RIS} depends only on the channels incident on that surface, and the channels from that surface to the users, $\bar{\mathbf{H}}_{\mathrm{TX}}^{(b)}$ and $\mathbf{H}_{\mathrm{RX},k}^{(b)}$, respectively, together with the direct channel term $\bar{\mathbf{H}}_k$ when the latter is retained. 
Therefore, the channels associated with the remaining \acp{BD-RIS} can be neglected in the passive update of $\mathbf{\Theta}^{(b)}$. }


\bibliographystyle{IEEEtran}
\bibliography{ref}

@techreport{3gpp36814,
  author      = {{3GPP}},
  title       = {{Further advancements for E-UTRA physical layer aspects}},
  institution = {{3rd Generation Partnership Project (3GPP)}},
  type        = {{Technical Report}},
  number      = {{TR 36.814}},
  version     = {{V9.2.0}},
  month       = mar,
  year        = {2017}
}

@misc{FidanovskiArx2025,
      title={Fractional Programming and Manifold Optimization for Reciprocal {BD-RIS} Scattering Matrix Design}, 
      author={Marko Fidanovski and Iv{\'a}n Alexander Morales Sandoval and Kuranage Roche Rayan Ranasinghe and Giuseppe Thadeu Freitas de Abreu and Emil Björnson and Bruno Clerckx},
      year={2025},
      eprint={2511.07683},
      archivePrefix={arXiv},
      primaryClass={eess.SP},
      url={https://arxiv.org/abs/2511.07683}, 
}

@ARTICLE{MirettiTSP2024,
  author={Miretti, Lorenzo and Cavalcante, Renato Lu{\'i}s Garrido and Bj{\"o}rnson, Emil and Sta{\'n}czak, S\l{}awomir},
  journal={IEEE Transactions on Signal Processing}, 
  title={{UL}-{DL} Duality for Cell-Free Massive {MIMO} With Per-{AP} Power and Information Constraints}, 
  year={2024},
  volume={72},
  number={},
  pages={1750-1765},
  doi={10.1109/TSP.2024.3376809}}

@ARTICLE{FangCL2024,
  author={Fang, Tianyu and Mao, Yijie},
  journal={IEEE Communications Letters}, 
  title={A Low-Complexity Beamforming Design for Beyond-Diagonal {RIS} Aided Multi-User Networks}, 
  year={2024},
  volume={28},
  number={1},
  pages={203-207},
  doi={10.1109/LCOMM.2023.3333411}}

@article{ShenTSP2018,
  author    = {Kaiming Shen and Wei Yu},
  title     = {Fractional Programming for Communication Systems—Part {I}: Power Control and Beamforming},
  journal   = {IEEE Transactions on Signal Processing},
  volume    = {66},
  number    = {10},
  pages     = {2616--2630},
  year      = {2018},
  month     = {May},
  doi       = {10.1109/TSP.2018.2819182},
  publisher = {IEEE}
}

@ARTICLE{Shen2TSP2018,
  author={Shen, Kaiming and Yu, Wei},
  journal={IEEE Transactions on Signal Processing}, 
  title={Fractional Programming for Communication Systems—Part {II}: Uplink Scheduling via Matching}, 
  year={2018},
  volume={66},
  number={10},
  pages={2631-2644},
  doi={10.1109/TSP.2018.2812748}}

@article{BjornsonSPM2022,
  author    = {E. Bj{\"o}rnson and H. Wymeersch and B. Matthiesen and P. Popovski and L. Sanguinetti and E. de Carvalho},
  title     = {Reconfigurable Intelligent Surfaces: A Signal Processing Perspective with Wireless Applications},
  journal   = {IEEE Signal Processing Magazine},
  volume    = {39},
  number    = {2},
  pages     = {135--158},
  year      = {2022},
  publisher = {IEEE},
  doi       = {10.1109/MSP.2021.3130549}
}

@article{LiuCST2021,
  author    = {Yuanwei Liu and Xiao Liu and Xidong Mu and Tianwei Hou and Jiaqi Xu and Marco Di Renzo and Naofal Al-Dhahir},
  title     = {Reconfigurable Intelligent Surfaces: Principles and Opportunities},
  journal   = {IEEE Commun. Surveys Tuts.},
  volume    = {23},
  number    = {3},
  pages     = {1546--1577},
  year      = {2021},
  publisher = {IEEE},
  doi       = {10.1109/COMST.2021.3077735}
}

@article{NeriniTWC2023,
  author    = {Matteo Nerini and Shanpu Shen and Bruno Clerckx},
  title     = {Closed-Form Global Optimization of Beyond Diagonal Reconfigurable Intelligent Surfaces},
  journal   = {IEEE Transactions on Wireless Communications},
  year      = {2023},
  doi       = {10.1109/TWC.2023.3307071},
  publisher = {IEEE}
}

@phdthesis{shen2018phd,
  author       = {Kaiming Shen},
  title        = {Fractional Programming for Communication System Design},
  school       = {Stanford University},
  year         = {2018},
  url          = {https://kaimingshen.github.io/doc/shen_thesis.pdf},
  note         = {Ph.D. dissertation, available online},
}

@ARTICLE{ZhouTWC2024,
  author={Zhou, Yuyan and Liu, Yang and Li, Hongyu and Wu, Qingqing and Shen, Shanpu and Clerckx, Bruno},
  journal={IEEE Transactions on Wireless Communications}, 
  title={Optimizing Power Consumption, Energy Efficiency, and Sum-Rate Using Beyond Diagonal {RIS}—A Unified Approach}, 
  year={2024},
  volume={23},
  number={7},
  pages={7423-7438},
  doi={10.1109/TWC.2023.3341262}}

@misc{FidanovskiTWC2026,
      title={Reciprocal Beyond-Diagonal Reconfigurable Intelligent Surface ({BD-RIS}): Scattering Matrix Design via Manifold Optimization}, 
      author={Marko Fidanovski and Iván Alexander Morales Sandoval and Hyeon Seok Rou and Giuseppe Thadeu Freitas de Abreu and Emil Bj{\"o}rnson},
      year={2026},
      eprint={2509.20246},
      archivePrefix={arXiv},
      primaryClass={eess.SP},
      url={https://arxiv.org/abs/2509.20246}, 
}

@ARTICLE{LiTSP2024,
  author={Li, Hongyu and Shen, Shanpu and Zhang, Yumeng and Clerckx, Bruno},
  journal={IEEE Transactions on Signal Processing}, 
  title={Channel Estimation and Beamforming for Beyond Diagonal Reconfigurable Intelligent Surfaces}, 
  year={2024},
  volume={72},
  number={},
  pages={3318-3332},
  doi={10.1109/TSP.2024.3424229}}

@INPROCEEDINGS{XuWCNC2025,
  author={Xu, Xiaolong and Ju, Ying and Li, Zhenghao and Hou, Xiangwang and Liu, Lei and Mumtaz, Shahid and Wu, Celimuge},
  booktitle={2025 IEEE Wireless Communications and Networking Conference (WCNC)}, 
  title={Beamforming Design for Multi-Sector {BD-RIS} Assisted {FL} with AirComp}, 
  year={2025},
  volume={},
  number={},
  pages={1-6},
  doi={10.1109/WCNC61545.2025.10978456}}

@article{LiTWC2022,
author = {Li, Hongyu and Shen, Shanpu and Clerckx, Bruno},
year = {2022},
month = {01},
pages = {1-1},
title = {Beyond Diagonal Reconfigurable Intelligent Surfaces: From Transmitting and Reflecting Modes to Single-, Group-, and Fully-Connected Architectures},
volume = {PP},
journal = {IEEE Transactions on Wireless Communications},
doi = {10.1109/TWC.2022.3210706}
}

@ARTICLE{LiJSAC2023,
  author={Li, Hongyu and Shen, Shanpu and Clerckx, Bruno},
  journal={IEEE Journal on Selected Areas in Communications}, 
  title={Beyond Diagonal Reconfigurable Intelligent Surfaces: A Multi-Sector Mode Enabling Highly Directional Full-Space Wireless Coverage}, 
  year={2023},
  volume={41},
  number={8},
  pages={2446-2460},
  doi={10.1109/JSAC.2023.3288251}}

@misc{GrantCVX2014,
  author       = {Michael Grant and Stephen Boyd},
  title        = {{CVX}: Matlab Software for Disciplined Convex Programming, version 2.1},
  howpublished = {\url{https://cvxr.com/cvx}},
  month        = mar,
  year         = 2014
}

@ARTICLE{LiuWCL2024,
  author={Liu, Zengrui and Liu, Yang and Shen, Shanpu and Wu, Qingqing and Shi, Qingjiang},
  journal={IEEE Wireless Communications Letters}, 
  title={Enhancing {ISAC} Network Throughput Using Beyond Diagonal {RIS}}, 
  year={2024},
  volume={13},
  number={6},
  pages={1670-1674},
  doi={10.1109/LWC.2024.3386155}}

@article{SandovalA2023,
author = {Sandoval, Iván and Ando, Kengo and Taghizadeh, Omid and Abreu, Giuseppe},
year = {2023},
month = {01},
pages = {1-1},
title = {Sum-Rate Maximization and Leakage Minimization for Multi-User Cell-Free Massive {MIMO} Systems},
volume = {PP},
journal = {IEEE Access},
doi = {10.1109/ACCESS.2023.3331767}
}

@ARTICLE{BjornsonSPM2024,
  author={Bj{\"o}rnson, Emil and Bengtsson, Mats and Ottersten, Bj{\"o}rn},
  journal={IEEE Signal Processing Magazine}, 
  title={Optimal Multiuser Transmit Beamforming: A Difficult Problem with a Simple Solution Structure [Lecture Notes]}, 
  year={2014},
  volume={31},
  number={4},
  pages={142-148},
  doi={10.1109/MSP.2014.2312183}}

@ARTICLE{GershmanSPM2010,
  author={Gershman, Alex B. and Sidiropoulos, Nicholas D. and Shahbazpanahi, Shahram and Bengtsson, Mats and Ottersten, Bjorn},
  journal={IEEE Signal Processing Magazine}, 
  title={Convex Optimization-Based Beamforming}, 
  year={2010},
  volume={27},
  number={3},
  pages={62-75},
  doi={10.1109/MSP.2010.936015}}

@book{GodaraCRCP2018,
  title={Handbook of antennas in wireless communications},
  author={Godara, Lal Chand},
  year={2018},
  publisher={CRC press}
}

@INPROCEEDINGS{ChenSPAWC2024,
  author={Chen, Kexin and Mao, Yijie},
  booktitle={2024 IEEE 25th International Workshop on Signal Processing Advances in Wireless Communications (SPAWC)}, 
  title={Transmitter Side Beyond-Diagonal {RIS} for mmWave Integrated Sensing and Communications}, 
  year={2024},
  volume={},
  number={},
  pages={951-955},
  doi={10.1109/SPAWC60668.2024.10693959}}

@INPROCEEDINGS{ShiIWCMC2016,
  author={Shi, Shuo and Zhu, Shida and Gu, Xuemai and Hu, Ruidong},
  booktitle={2016 International Wireless Communications and Mobile Computing Conference (IWCMC)}, 
  title={Extendable carrier synchronization for distributed beamforming in wireless sensor networks}, 
  year={2016},
  volume={},
  number={},
  pages={298-303},
  doi={10.1109/IWCMC.2016.7577074}}

@article{GhaniPONE2016,
author = {Ghani, Anwar and Naqvi, Husnain and Ramzan, Muhammad Sher and Khattak, Muazzam and Khan, Imran and Irshad, Azeem},
year = {2016},
month = {07},
pages = {e0159069},
title = {Spread Spectrum Based Energy Efficient Collaborative Communication in Wireless Sensor Networks},
volume = {11},
journal = {PLOS ONE},
doi = {10.1371/journal.pone.0159069}
}

@article{PearceER2020,
title = {Limiting liability with positioning to minimize negative health effects of cellular phone towers},
journal = {Environmental Research},
volume = {181},
pages = {108845},
year = {2020},
issn = {0013-9351},
doi = {https://doi.org/10.1016/j.envres.2019.108845},
url = {https://www.sciencedirect.com/science/article/pii/S0013935119306425},
author = {J.M. Pearce}
}

@article{RodaESCP2014,
title = {Mobile phone infrastructure regulation in Europe: Scientific challenges and human rights protection},
journal = {Environmental Science and Policy},
volume = {37},
pages = {204-214},
year = {2014},
issn = {1462-9011},
doi = {https://doi.org/10.1016/j.envsci.2013.09.009},
url = {https://www.sciencedirect.com/science/article/pii/S146290111300186X},
author = {Claudia Roda and Susan Perry}
}

@article{MaTWC2023,
  title={Cooperative beamforming for {RIS}-aided cell-free massive {MIMO} networks},
  author={Ma, Xinying and Zhang, Deyou and Xiao, Ming and Huang, Chongwen and Chen, Zhi},
  journal={IEEE Transactions on Wireless Communications},
  volume={22},
  number={11},
  pages={7243--7258},
  year={2023},
  publisher={IEEE}
}

@article{LiuJSTSP2025,
  title={Distributed Distortion-Aware Beamforming Designs for Cell-Free {mMIMO} Systems},
  author={Liu, Mengzhen and Li, Ming and Liu, Rang and Liu, Qian},
  journal={IEEE Journal of Selected Topics in Signal Processing},
  year={2025},
  publisher={IEEE}
}

@article{BjornsonDSP2019,
title = {Massive {MIMO} is a reality—What is next?: Five promising research directions for antenna arrays},
journal = {Digital Signal Processing},
volume = {94},
pages = {3-20},
year = {2019},
note = {Special Issue on Source Localization in Massive MIMO},
issn = {1051-2004},
doi = {https://doi.org/10.1016/j.dsp.2019.06.007},
url = {https://www.sciencedirect.com/science/article/pii/S1051200419300776},
author = {Emil Bj{\"o}rnson and Luca Sanguinetti and Henk Wymeersch and Jakob Hoydis and Thomas L. Marzetta}
}

@misc{KatsanosArx2026,
      title={Decentralized Cooperative Beamforming for {BDRIS}-Assisted Cell-Free {MIMO} {OFDM} Systems}, 
      author={Konstantinos D. Katsanos and George C. Alexandropoulos},
      year={2026},
      eprint={2601.13201},
      archivePrefix={arXiv},
      primaryClass={eess.SP},
      url={https://arxiv.org/abs/2601.13201}, 
}

@misc{WuArx2024,
      title={Optimization of Beyond Diagonal {RIS}: A Universal Framework Applicable to Arbitrary Architectures}, 
      author={Zheyu Wu and Bruno Clerckx},
      year={2024},
      eprint={2412.15965},
      archivePrefix={arXiv},
      primaryClass={eess.SP},
      url={https://arxiv.org/abs/2412.15965}, 
}

\end{document}